\documentclass[aps,pra,showpacs,twocolumn,superscriptaddress,10pt]{revtex4}
\usepackage{xy,amsmath,amssymb,amsthm,graphics,graphicx,bm,subfigure,epstopdf,multirow,tabularx,natbib,url}
\usepackage[colorlinks,linkcolor=blue]{hyperref}
\input{Qcircuit}

\newcommand{\be}{\begin{equation}}
\newcommand{\ee}{\end{equation}}
\newcommand{\ben}{\begin{align}}
\newcommand{\een}{\end{align}}
\newcommand{\bes}{\begin{subequations}}
\newcommand{\ees}{\end{subequations}}
\newcommand{\bF}{\begin{figure}}
\newcommand{\eF}{\end{figure}}
\newcommand{\bW}{\begin{widetext}}
\newcommand{\eW}{\end{widetext}}
\newcommand{\dg}{\dagger}

\def\ket#1{ | #1 \rangle}
\def\bra#1{{\langle #1 |  }}

\newcommand{\proj}[1]{\mbox{$|#1\rangle \!\langle #1 |$}}

\newcommand{\Tr}[1]{\textrm{Tr}\left[#1\right]}

\begin{document}

\title{Strategies for enhancing quantum entanglement by local photon subtraction}

\author{Tim J. Bartley}
\email{t.bartley1@physics.ox.ac.uk}
\affiliation{Clarendon Laboratory, Department of Physics, University of Oxford, OX1 3PU, United Kingdom}

\author{Philip J. D. Crowley}
\affiliation{Clarendon Laboratory, Department of Physics, University of Oxford, OX1 3PU, United Kingdom}

\author{Animesh Datta}
\affiliation{Clarendon Laboratory, Department of Physics, University of Oxford, OX1 3PU, United Kingdom}

\author{Joshua Nunn}
\affiliation{Clarendon Laboratory, Department of Physics, University of Oxford, OX1 3PU, United Kingdom}

\author{Lijian Zhang}
\affiliation{Clarendon Laboratory, Department of Physics, University of Oxford, OX1 3PU, United Kingdom}
\affiliation{Max-Planck Institute for Structural Dynamics, University of Hamburg, 22607 Hamburg, Germany}

\author{Ian Walmsley}
\affiliation{Clarendon Laboratory, Department of Physics, University of Oxford, OX1 3PU, United Kingdom}

\date{\today}

\begin{abstract}
Subtracting photons from a two-mode squeezed state is a well-known method to increase entanglement. We analyse different strategies of local photon subtraction from a two-mode squeezed state in terms of entanglement gain and success probability. We develop a general framework that incorporates imperfections and losses in all stages of the process: before, during, and after subtraction. By combining all three effects into a single efficiency parameter, we provide analytical and numerical results for subtraction strategies using photon-number-resolving and threshold detectors. We compare the entanglement gain afforded by symmetric and asymmetric subtraction scenarios across the two modes. For a given amount of loss, we identify an optimised set of parameters, such as initial squeezing and subtraction beam splitter transmissivity, that maximise the entanglement gain rate. We identify regimes for which asymmetric subtraction of different Fock states on the two modes outperforms symmetric strategies. In the lossless limit, subtracting a single photon from one mode always produces the highest entanglement gain rate. In the lossy case, the optimal strategy depends strongly on the losses on each mode individually, such that there is no general optimal strategy. Rather, taking losses on each mode as the only input parameters, we can identify the optimal subtraction strategy and required beam splitter transmissivities and initial squeezing parameter. Finally, we discuss the implications of our results for the distillation of continuous-variable quantum entanglement.
\end{abstract}


\maketitle

\section{Introduction}\label{sec:Intro}
Efficient distribution of entanglement between distant parties is fundamental to most quantum communication protocols. However, entanglement is fragile and suffers from decoherence, which is detrimental to the performance of any communication protocol upon which it relies. The ability to increase the entanglement between communicating parties is therefore vital, and further, practical considerations dictate that this should be achieved through only local operations and classical communication (LOCC). While entanglement cannot increase \textit{on average} under LOCC, a probabilistic protocol can be employed to increase the entanglement of a subset of states. This is the basis of entanglement distillation: extracting a small ensemble of more strongly entangled states from a larger ensemble of weakly entangled states~\cite{Bennett:1996bh}. 

In the discrete variable regime, entanglement distillation has been achieved using photonic qubits~\cite{Kwiat_01}. In the continuous-variable (CV) regime, the situation is more involved. Most common CV states and operations are Gaussian in nature. However, there exists a no-go theorem which states that one cannot distill entanglement from Gaussian states by Gaussian operations alone~\cite{Eisert:2002zr,Fiurasek_prl02,Giedke_pra02}. Gaussian operations are those with Hamiltonians which are (at most) quadratic in the ladder operators $\hat{a},~\hat{a}^\dag$, comprising the basic tools of quantum optics including beam splitters, phase shifters, squeezers and homodyne detection. A number of protocols to increase entanglement in CV systems have been proposed~\cite{Opatrny_kw00,Duan2000,Fiurasek03,Browne:2003fk,Campbell_prl12,campbell_arx12}, elements of which have been implemented~\cite{Hage_SDFFS08,Dong_LHMFLA08,Hage_SDFS10,Dong_LHMFLA10,Sasaki_distillation}. 


Photon subtraction was initially proposed by Opatrn{\'y} \textit{et al.}\ to increase the efficacy of a teleportation protocol~\cite{Opatrny_kw00}. Since then, several studies have looked at photon subtraction in more detail. Cochrane \textit{et~al.} investigated subtracting and detecting $n$ photons simultaneously from the modes of a two-mode squeezed state~\cite{Cochrane_rm02}. Olivares  \textit{et~al.} studied the use of on-off (single-photon threshold) detectors to measure at least one photon subtracted from both modes coincidentally, again in terms of the improvement of a teleportation protocol~\cite{Olivares_pr03}. Kitagawa \textit{et~al.} provided a detailed numerical analysis of two-mode subtraction by on-off detectors in terms of the explicit change in entanglement and compared this with the operational measures used previously in the literature~\cite{Kitagawa_tsc06}. This work was  built on by Zhang and van Loock~\cite{Zhang_vL11} in which analytical results for perfect symmetric subtraction using photon number resolving detectors and on-off detectors were derived. More recently, Navarrete-Benlloch \textit{et~al.} extended the analysis to asymmetric subtraction and quantified the non-Gaussianity of the operations~\cite{Navarrete12}. Photon subtraction (and addition) is discussed more generally in terms of quantum state engineering in the review by Kim~\cite{Kim_jpb08}, and in terms of non-Gaussian entanglement quantification in~\cite{Adesso_pra09}. Experimentally, both nonlocal~\cite{Ourjoumtsev_prl07} and local~\cite{Sasaki_distillation} photon subtraction from two-mode squeezed states have been demonstrated.

In this paper, we extensively investigate the best entanglement enhancement strategy in a realistic experimental scenario. We present practical figures of merit based on the entanglement gain and success probability of photon subtraction protocols. Using the log-negativity~\cite{Plenio05} as an entanglement metric allows us to quantify the entanglement of mixed states caused by losses. Indeed, we consider six independent loss parameters and allow for detecting different numbers of photons subtracted from each mode, which we refer to as asymmetric subtraction. In addition, our model considers both on-off and photon-number resolving detectors to measure the subtracted photons. Our results can be applied directly to realistic experiments, as well as providing a framework in which to study other entanglement-enhancing strategies.

The article is organised as follows: in Sec.~\ref{sec:EntMet}, we define the figures of merit by which protocols that increase entanglement may be measured. In Sec.~\ref{sec:StateEvolution}, we derive the state evolution of photon subtraction from a two-mode squeezed state (TMSS) and we provide an analytical form for the probabilities corresponding to different detection strategies. We use the evolved state to analyse the gain in entanglement by different subtraction strategies: using photon-number resolving detectors (PNRDs) in Sec.~\ref{sec:PNRD}, and threshold detectors (such as avalanche photodiodes --- APDs) in Sec.~\ref{sec:APD}. For both type of detectors (PNRDs in Sec.~\ref{sec:losses} and APDs in Sec.~\ref{sec:APDloss}), we numerically analyse symmetric and asymmetric subtraction in the presence of loss occurring before, during, and after subtraction, and show how these different parameters affect the success of the protocol. 
We also list the main conclusions drawn from our work in Sec.~\ref{sec:conc}.



\section{Entanglement, gain and rate}
\label{sec:EntMet}
Entanglement in this system can be captured conveniently by the positive partial transpose (PPT) criterion. If the partial transpose $\rho^{T_A}$  of a density matrix $\rho$ has negative eigenvalues, then $\rho$  must be entangled~\cite{Peres96,Horodecki1996}. The sum of the absolute values of the negative eigenvalues of $\rho^{T_A}$  is defined as the negativity $N\left(\rho\right)$. The entanglement measure we use in this paper is the log-negativity~\cite{Plenio05}, defined
\be
\label{eqn:LogNeg}
E_N\left(\rho\right)=\log_2\left[1+2N\left(\rho\right)\right]=\log_2\left|\left|\rho^{T_A}\right|\right|_1~,
\ee
where $\left|\left|X\right|\right|_1=\Tr{\sqrt{X^\dag X}}$ denotes the trace norm of $X$. 

The maximally entangled CV state, for a fixed energy, is the two-mode squeezed state (TMSS), as generated during parametric down-conversion (PDC) or equivalently by interfering two single-mode squeezed vacua in phase at a 50:50 beam splitter~\cite{Braunstein:2005ly}. This state can be written in the Fock basis as
\begin{align}\label{eqn:TMSS}
\ket{\psi_{\mathrm{TMSS}}} &=\sqrt{1-\lambda^2}\sum_{n=0}^\infty\lambda^n \ket{n,n}_{AA^\prime}~,\\
&=\sqrt{1-\lambda^2}\sum_{n=0}^\infty \frac{\lambda^n}{n!}a^{\dag n}a^{\prime \dag n}\ket{0,0}_{AA^\prime}~,
\end{align}
which describes $n$ pairs of photons in modes $A$ and $A^\prime$ for a given squeezing parameter $\lambda\in [0,1)$ and we define $\ket{n,m}_{AA^\prime}=\ket{n}_{A}\otimes \ket{m}_{A^\prime}$. 

The density matrix of this state is
\be
\rho_\textrm{TMSS}=\left(1-\lambda^2\right)\sum_{n=0}^\infty\sum_{m=0}^\infty\lambda^n\lambda^m\ket{n}{_A}\bra{m}\otimes\ket{n}{_{A^\prime}}\bra{m}~,
\ee
the partial transpose (with respect to mode $A$) of which is
\be
\rho^{T_A}_\textrm{TMSS}=\left(1-\lambda^2\right)\sum_{n=0}^\infty\sum_{m=0}^\infty\lambda^n\lambda^m\ket{m}{_A}\bra{n}\otimes\ket{n}{_{A^\prime}}\bra{m}~.
\ee
Taking the trace norm yields
\begin{align}\nonumber
\left|\left|\rho_\textrm{TMSS}^{T_A}\right|\right|_1&=\left(1-\lambda^2\right)\Tr{\sum_{n=0}^\infty\lambda^n\proj{n}}\Tr{\sum_{m=0}^\infty\lambda^m\proj{m}}\\
&=\frac{\left(1-\lambda^2\right)}{\left(1-\lambda\right)^2}~,
\end{align}
therefore the TMSS defined in Eqn.~(\ref{eqn:TMSS}) has log-negativity
\be\label{eqn:EntInit}
E_N\left(\rho_{\mathrm{TMSS}}\right)=\log_2\left(\frac{1+\lambda}{1-\lambda}\right)~,
\ee
which is the benchmark against which changes in entanglement will be measured.

From this initial state $\rho_\textrm{TMSS}$, a subtraction step $\mathfrak{s}$ produces a state $\rho_\mathfrak{s}$ with entanglement $E_N\left(\rho_\mathfrak{s}\right)$. We define the \textit{gain} in entanglement $G\left(\rho_\mathfrak{s}\right)$ as the difference between entanglement after distillation and that of the initial state, normalised to the initial entanglement, \emph{i.e.}\
\be
\label{eqn:EntGain}
G\left(\rho_\mathfrak{s}\right)=\frac{E_N\left(\rho_\mathfrak{s}\right)}{E_N\left(\rho_\textrm{TMSS}\right)}-1~,
\ee
such that $G\left(\rho_\mathfrak{s}\right) > 0$ if the subtraction step increases entanglement. 

The probability of an entanglement enhancing step must be less than unity, since entanglement cannot be increased on average under LOCC~\cite{Nielsen_99}. It is calculated by 
\be\label{eqn:EntProb}
P\left(\rho_\mathfrak{s}\right)=\Tr{\mathfrak{s}\left(\rho\right)}~,
\ee
where $\mathfrak{s}\left(\rho\right)$ is the unnormalised density matrix following a subtraction operation. The final density matrix $\rho_\mathfrak{s}$ is found by normalising $\mathfrak{s}\left(\rho\right)$ thus
\be\label{eqn:Normalise}
\rho_\mathfrak{s} = \frac{\mathfrak{s}\left(\rho\right)}{\Tr{\mathfrak{s}\left(\rho\right)}}~.
\ee

The aim of an entanglement enhancement protocol is to increase entanglement from some initial value. It is therefore desirable not only for subtraction to produce a high gain, but also to do so at a high rate, \emph{i.e.}\ that the probability of a subtraction event is as high as possible. 
In general these conditions cannot be met independently, therefore there exists an optimum to be found based on the parameters of a given implementation. We therefore define the \textit{entanglement gain rate} (frequently shortened to ``rate'' in the remainder of the paper) $\Gamma\left(\rho_\mathfrak{s}\right)$ as the gain afforded by a distillation step $G\left(\rho_\mathfrak{s}\right)$ multiplied by its likelihood
\be
\label{eqn:EntEff}
\Gamma\left(\rho_\mathfrak{s}\right)=P\left(\rho_\mathfrak{s}\right)G\left(\rho_\mathfrak{s}\right)~.
\ee
By post-selecting on successful subtraction events, the entanglement of this sub-ensemble can be increased. The increase will depend on the parameters of the subtraction step employed, and can be maximised to determine the values of those parameters which yield the highest rate.


\section{The protocol}\label{sec:StateEvolution}
\bF
\includegraphics[width=0.4\textwidth]{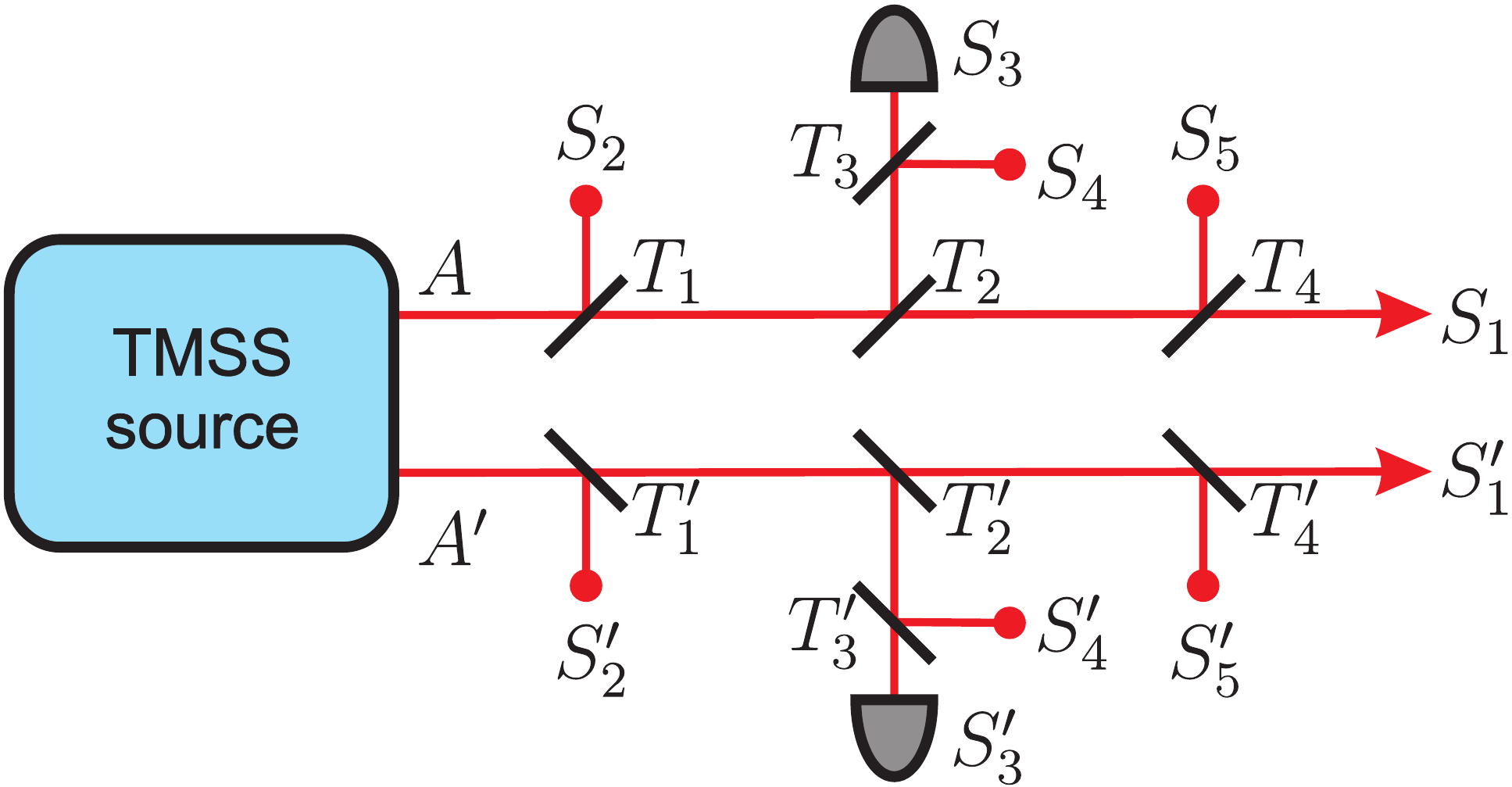}
\caption{(Color online) A two-mode squeezed state (TMSS) in a photon-subtraction setup. A TMSS initially occupies modes $A,A^\prime$. Following the array of beam splitters (with transmissivity $T_i,T^{\prime}_i$), we seek the entanglement in modes $S_1,S_1^\prime$ following detection in modes $S_3,S_3^\prime$. Losses are modelled by photons in modes before ($S_2,S_2^\prime$), during ($S_4,S_4^\prime$) and after ($S_5,S_5^\prime$) detection. }\label{fig:Subtr_setup}
\eF

To effect subtraction and account for loss, we consider an array of $8$ beam splitters, each with transmissivity $T_i$ and reflectivity $R_i=\sqrt{1-T^2_i}$, acting on the TMSS as shown in Fig.~\ref{fig:Subtr_setup}. The state initially occupies modes $A,A'$. The beam splitters $T_2,T_2^\prime$ effect subtraction, while the other beam splitters $T_1,T_3,T_4$ model losses before subtraction, during detection and after subtraction, respectively (and similarly for the primed counterparts). As such, the state vector describing the combined state of the input modes may be written $\ket{\Psi_{\textrm{in}}}= \ket{\psi_\textrm{TMSS}}\otimes\ket{0}^{\otimes 8}$. The evolution of the input modes is governed by the unitary transformation 
$U\oplus U^\prime$. The unitaries $U$ and $U'$ actually denote the orthogonal (rotation) matrices corresponding to symplectic transformations in the Heisenberg picture, and operates on the modes as labelled in Fig.~\ref{fig:Subtr_setup} and circuit diagram~(\ref{eqn:qcircuit}).
The elements of $U,U^{\prime}$ depend on the arrangement of beam splitters and how they couple their respective modes. For the array of beam splitters shown in Fig.~\ref{fig:Subtr_setup}, $U$ may be written as
\be
U\!\!=\!\!\left(\begin{array}{ccccc}
T_1T_2T_4 & -R_1T_2T_4 &-R_2T_4 & 0 & -R_4\\
R_1 & T_1 & 0 & 0  & 0\\
T_1R_2T_3 & -R_1R_2T_3 & T_2T_3 & -R_3&0\\
T_1R_2R_3 & -R_1R_2R_3 & T_2R_3 & T_3&0\\
T_1T_2R_4 & -R_1T_2R_4 & -R_2R_4 & 0 & T_4
\end{array}\right)~,
\ee
and similarly for $U^\prime$, with all symbols replaced by their primed counterparts.
Since the only non-vacuum input modes are $A,A^\prime$, the output modes depend only on the first column of each component unitary $U,U^\prime$. Indeed, we may write the output state vector in the Fock basis as
\begin{align}
\label{eqn:state1}
\ket{\Psi_\textrm{out}}=&\sqrt{1-\lambda^2}\sum_{n=0}^\infty\frac{\lambda^n}{n!}\left(\sum_{m=1}^5\sigma_{1,m}s^\dg_m\right)^n \nonumber \\
& \times \left(\sum^{5}_{m^\prime=1}\sigma_{1,m^\prime}^\prime s^{\prime\dg}_{m^\prime}\right)^n \ket{0}~,
\end{align}
where $\sigma_{1,m}$ is the $(1,m)$-th element of the unitary $U$ and $s^\dag_m$ is the creation operator for the mode $m$. 
We are interested in the entanglement between the modes $S_1,S_1^\prime$, occupied by $e,e^\prime$ photons respectively, dependent on detecting $d,d^\prime$ photons in modes $S_3,S_3^\prime$, respectively. The modes $S_2,S_4,S_5,S_2^\prime,S_4^\prime,S_5^\prime$ contain $l,l^\prime$ photons lost to the environment. These loss modes can be combined, simplifying our problem into the following circuit diagram
\be\label{eqn:qcircuit}
\Qcircuit @C=1em @R=0.7em {
&&& \lstick{\ket{0}^{\otimes 3}} & \multigate{2}{U} & \ustick{\gamma}\qw & \rstick{\ket{l}^{\otimes 3}} \qw &&\\
&&& \lstick{\ket{0}} & \ghost{U}& \ustick{\beta}\qw & \measureD{d} &&\\
&&& \lstick{~\ket{n}}& \ghost{U} & \ustick{\alpha}\qw &\rstick{\ket{e}~} \qw \gategroup{3}{2}{5}{2}{0.2em}{\{} \gategroup{3}{8}{5}{8}{0.2em}{\}}&&\\
&\lstick{\ket{\psi_\textrm{TMSS}}~} & & & & & &  \rstick{~\mathfrak{s}\left(\rho\right)}&&\\
&&& \lstick{~\ket{n}}& \multigate{2}{U^\prime} & \ustick{\alpha^\prime}\qw &\rstick{\ket{e^\prime}~} \qw &&\\
&&& \lstick{\ket{0}} & \ghost{U^\prime}& \ustick{\beta^\prime}\qw &\measureD{d^\prime} &&\\
&&& \lstick{\ket{0}^{\otimes 3}} & \ghost{U^\prime} & \ustick{\gamma^\prime}\qw &\rstick{\ket{l^\prime}^{\otimes 3}} \qw &&\\
}
\ee
\\
The photons in two input modes of the TMSS are divided into three output modes each: entangled, detected and lost, and we assign parameters $\alpha,\alpha^\prime$, $\beta, \beta^\prime$ and $\gamma,\gamma^\prime$ to be the fraction of photons in each mode, respectively. These may be written in terms of the components of the unitary $U$ thus
\bes
\label{eqn:coeffs}
\begin{align}
\alpha^2&\equiv\sigma^2_{1,1}=T^2_1T^2_2T^2_4~,\label{eqn:alpha}\\
\beta^2&\equiv\sigma^2_{1,3}=T^2_1R^2_2T^2_3~,\label{eqn:Beta}\\
\gamma^2&\equiv\sum_{m=2,4,5}\sigma^2_{1,m}=R^2_1+T^2_1\left(R^2_2R^2_3+T^2_2R^2_4\right)~,\label{eqn:Gamma}
\end{align}
\ees
where the coefficients satisfy $\alpha^2+\beta^2+\gamma^2=1$, as defined by the unitarity condition 
(and similarly for their primed counterparts). Substituting Eqns.~(\ref{eqn:coeffs}) into Eqn.~(\ref{eqn:state1}),
yields
\be
\ket{\Psi_\textrm{out}}=\sqrt{1-\lambda^2}\sum_{n=0}^\infty\lambda^n\!\sum_{\mathcal{S}}c_{e,d,l}c^{\prime}_{e^\prime,d^\prime,l^\prime}\ket{e,e^\prime}\ket{d,d^\prime}\ket{l,l^\prime}~,
\ee
where the summation is over $\mathcal{S} = \{e,e^\prime,d,d^\prime,l,l^\prime~\textrm{s.t.}~e+d+l=e^\prime+d^\prime+ l^\prime=n\}$, and where we have defined 
\be
c_{e,d,l}=\sqrt{\left(\begin{array}{c}n\\e,d,l\end{array}\right)}\alpha^e\beta^d\gamma^l~,
\ee
with the multinomial coefficient~\cite{DLMF} $\left(\begin{array}{c}n\\e,d,l\end{array}\right)$ (and similarly for the primed quantities). On tracing out the loss and detected modes, we obtain the (mixed) entangled state $\rho$ across modes $S_1,S_1^\prime$ as
\be\label{eqn:Rho}
\rho=(1-\lambda^2) \sum_{n,\tilde{n}=0}^\infty\sum_{d,d^\prime}^{d_\textrm{max},d_\textrm{max}^\prime} \sum_{l,l^\prime}^{l_\textrm{max}}c_{n, \tilde{n}, d,d^\prime, l,l^\prime}\ket{e}_{S_1}\bra{\tilde{e}}\otimes\ket{e^\prime}_{S_1^\prime}\bra{\tilde{e}^\prime}~,
\ee
where $d_\textrm{max}=\min\left(n,\tilde{n}\right)-l$, $l_\textrm{max}=\min\left(n,\tilde{n}\right)$, and $c_{n,\tilde{n},d,d^\prime,l,l^\prime}$ are subject to the constraints
$e+d+l=e^{\prime}+d^{\prime}+l^{\prime}=n$, $\tilde{e}+d+l=\tilde{e}^{\prime}+d^{\prime}+l^{\prime}=\tilde{n}$, such that
\be
c_{n,\tilde{n},d,d^\prime, l,l^\prime}=\frac{\lambda^{n+\tilde{n}}n!\tilde{n}!\alpha^{e+\tilde{e}}\alpha^{\prime e^\prime+\tilde{e}^\prime}\beta^{2d}\beta^{\prime 2d^\prime}\gamma^{2l}\gamma^{\prime 2l^\prime}}{d!d^\prime!l!l^\prime!\sqrt{e!e^\prime!\tilde{e}!\tilde{e}^\prime!}}~.
\ee
As the mode structure is now clear, we will suppress the mode labels $S_1,S_1^\prime$ in further discussions for compactness.

\subsection{Subtraction}
Subtraction is effected by changing the limits of the summation over $d,d^\prime$ in Eqn.~(\ref{eqn:Rho}), leading to an unnormalised subtracted state $\mathfrak{s}\left(\rho\right)$. The exact form of this state depends on the subtraction strategy employed and is derived in detail in the relevant sections below. 
To calculate the entanglement we seek the sum of the negative eigenvalues of the partial transpose of the normalised state. Using Eqns.~(\ref{eqn:Normalise}) and ~(\ref{eqn:LogNeg}), we therefore seek
\be 
E_n\left(\rho_\mathfrak{s}\right)=\log_2\left\{\frac{\left|\left|\mathfrak{s}\left(\rho\right)^{T_{S_1}}\right|\right|_1}{\Tr{\mathfrak{s}\left(\rho\right)}}\right\}~,
\ee
where $\mathfrak{s}\left(\rho\right)^{T_{S_1}}$ is the partially transposed unnormalised state with respect to mode $S_1$.
Following the analysis in Refs.~\cite{Kitagawa_tsc06,Zhang_vL11}, we can write $\mathfrak{s}\left(\rho\right)^{T_{S_1}}$ in block diagonal form, and we change variables such that
\be
n\equiv i+l+d,~\tilde{n}\equiv j+l+d,~l^\prime \equiv i+j-K+l+d-d^\prime~,
\ee
where the indices $i,j$ denote the rows and columns of the $K^{\mathrm{th}}$ block matrix of dimension $K+1$. Thus $\mathfrak{s}\left(\rho\right)^{T_{S_1}}$ is written explicitly as
\be
\mathfrak{s}\left(\rho\right)^{T_{S_1}}=\left(1-\lambda^2\right)\bigoplus_{K=0}^\infty\sum_{i,j=0}^KC_{i,j}^{\left(K\right)}\ket{j}\bra{i}\otimes\ket{K-i}\bra{K-j}~,
\ee
with coefficients
\begin{widetext}
\be
\label{eqn:Ckij}
C_{i,j}^{\left(K\right)}=\sum_{d,d^\prime=t,t^\prime}^{t_{\textrm{max}},t^\prime_{\textrm{max}}} \sum_{l=l_0}^\infty \frac{\lambda^{i+j+2\left(l+d\right)} \alpha^{i+j} \alpha^{\prime 2K-i-j} \beta^{2d} \beta^{\prime2d^\prime} \gamma^{2l} \gamma^{\prime 2\left(i+j-K+l+d-d^\prime\right)} \left(i+l+d\right)!\left(j+l+d\right)!} {l!\left(l+i+j+d-K-d^\prime\right)!d!d^\prime! \sqrt{i!j!\left(K-j\right)! \left(K-i\right)!}}~,
\ee
\end{widetext}
where $l_0=\max{\left\{0,K+d^\prime-d-i-j\right\}}$ and $t,t^\prime,t_\textrm{max},t^\prime_\textrm{max}$ depend on the type of detector employed. By inspection the matrices $\mathbf{C}^{\left(K\right)}=\left[C_{i,j}^{\left(K\right)}\right]_{i=0,\ldots,K;j=0,\ldots,K}$ are symmetric, and they are also persymmetric (and therefore centrosymmetric) when the primed parameters are equal to their unprimed counterparts and the range of the summations over $d$ and $d^\prime$ are equal~\cite{Zhang_vL11}. Eqn.~(\ref{eqn:Ckij}) is central to the derivation of all subsequent analytic results.

The coefficients $C_{i,j}^{\left(K\right)}$ lead to a normalised state if $d,d^\prime$ are summed over entirely, \emph{i.e.}\ the limits on the summation over $d,d^\prime$ are $t=t^\prime=0,~t_\textrm{max}=t_\textrm{max}^\prime=\infty$ respectively. The (unnormalised) state $\mathfrak{s}\left(\rho\right)$ following a subtraction event is found by placing limits on the $d,d^\prime$ summation, yielding coefficients $\tilde{C}_{i,j}^{\left(K\right)}=C_{i,j}^{\left(K\right)}\left(t,t^\prime,t_\textrm{max},t_\textrm{max}^\prime\right)$, dependent on the strategy employed. For threshold detectors that click on receipt of a minimum number of photons, the sum runs from the threshold value $t$ to $t_\textrm{max}=\infty$ (for instance, single-photon avalanche detectors have $t=1$). For photon-number-resolving-detectors (PNRDs) operating within their resolution regime, the summation disappears since $t_\textrm{max}=t$.

\section{Subtraction using PNRDs}\label{sec:PNRD}
\begin{figure*}[thb]
\subfigure[]{\includegraphics[width=0.3\textwidth]{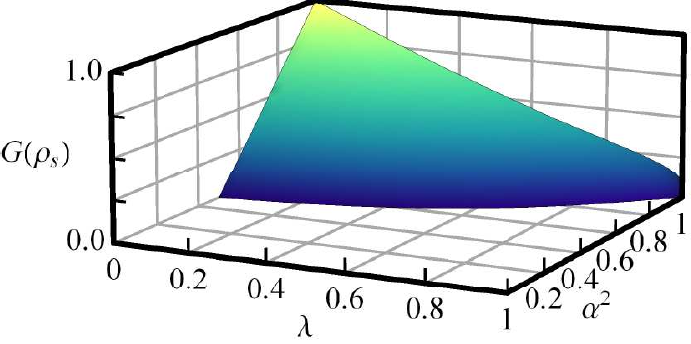}\label{fig:Gain}}\qquad
\subfigure[]{\includegraphics[width=0.3\textwidth]{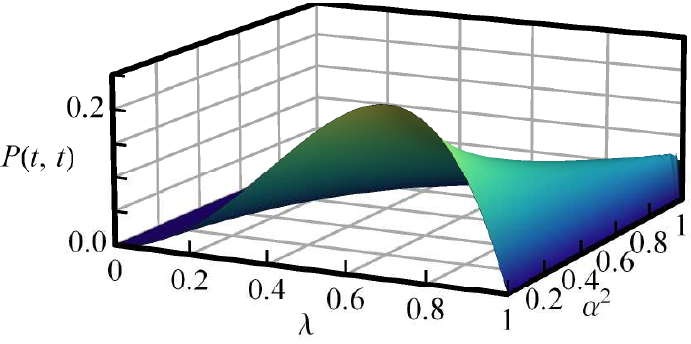}\label{fig:Probability}}\qquad
\subfigure[]{\includegraphics[width=0.3\textwidth]{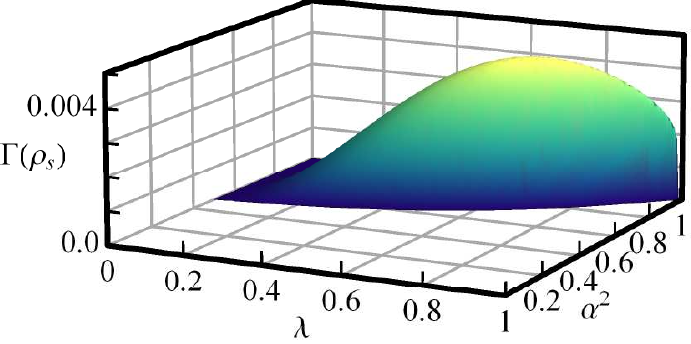}\label{fig:Rate}}
\caption{(Color online) Lossless symmetric subtraction of a single photon using perfect photon-number resolving detectors. Entanglement gain~\subref{fig:Gain}, distillation probability~\subref{fig:Probability} and rate~\subref{fig:Rate} are shown for $\lambda \in [0,1)$ and $\alpha^2 \in [0,1)$.}\label{fig:SymmSub}
\end{figure*}
We will first briefly consider the case when subtracted photons are measured using perfect PNRDs. This amounts to removing the sums over $d,d^\prime$ in Eqn.~(\ref{eqn:Ckij}). 

\subsection{Probability of photon subtraction}
The probability $P\left(t,t^\prime\right)=P\left(\rho_\mathfrak{s}\right)$ of 
detecting $t=t_\textrm{max}$, $t^\prime=t^\prime_\textrm{max}$ photons, respectively, is given by the trace of the unnormalised state $\mathfrak{s}\left(\rho\right)$ after a subtraction event. Detecting $t,t^\prime$ subtracted photons by PNRDs projects onto modes $S_1,S_1^\prime$ the unnormalised state 
\be
\mathfrak{s}\left(\rho\right)=\left(1-\lambda^2\right)\bigoplus_{K=0}^\infty\sum_{i,j=0}^K\tilde{C}_{i,j}^{\left(K\right)}\ket{i}_{E}\bra{j}\otimes\ket{K-i}_{E^\prime}\bra{K-j}~,
\ee
where the coefficients $\tilde{C}_{i,j}^{\left(K\right)}=C_{i,j}^{\left(K\right)}\left(t,t^\prime,t,t^\prime\right)$ are identical to those in Eqn.~(\ref{eqn:Ckij}) for $t_\textrm{max}=t,t_\textrm{max}^\prime=t^\prime$ and therefore the summation over $d$ is dropped.
This has probability
\bW
\begin{equation}\label{eqn:TrRho}
P\left(t,t^\prime\right) = \Tr{\mathfrak{s}\left(\rho\right)} = \frac{\left(1-\lambda^2\right) \lambda^{2t^\prime} \beta^{\prime2t^\prime} \beta^{2t} \left( \alpha^2 + \gamma^2 \right)^{t^\prime-t} }{ \left[1-\left( \alpha^2+\gamma^2 \right) \left( \alpha^{\prime2}+\gamma^{\prime2}  \right) \lambda^2 \right]^{t^\prime+1} } P_{t}^{(t^\prime-t,0 )} \left[ \frac{1+\left( \alpha^2+\gamma^2 \right) \left( \alpha^{\prime2}+\gamma^{\prime2} \right) \lambda^2 }{1-\left( \alpha^2+\gamma^2 \right) \left( \alpha^{\prime2}+\gamma^{\prime2}  \right) \lambda^2 } \right]~,
\end{equation}
\eW
where $P_n^{(a,b)}[z]$ denotes the $n^{\mathrm{th}}$-order Jacobi polynomial. We have assumed, without loss of generality, that $t\leq t^\prime$; owing to the symmetry of the problem a corresponding equation for $t \geq t^\prime$ can be found by swapping all of the primed values with their unprimed counterparts. This expression is valid if $t,t^\prime$ are known, \emph{i.e.}\ in the case where detection resolves photon number. This may be extended to threshold detectors by summing over $t,t^\prime$ from the threshold $\tilde{t},\tilde{t}^\prime$ (typically unity) to $t_\textrm{max},t^\prime_\textrm{max}$. We deal with this scenario in greater detail in Sec.~\ref{sec:APD}.

\subsection{Lossless symmetric subtraction}\label{sec:PNRDSymmetric}
\begin{figure}[tbh!]
\centering
\subfigure[]{\includegraphics[width=0.35\textwidth]{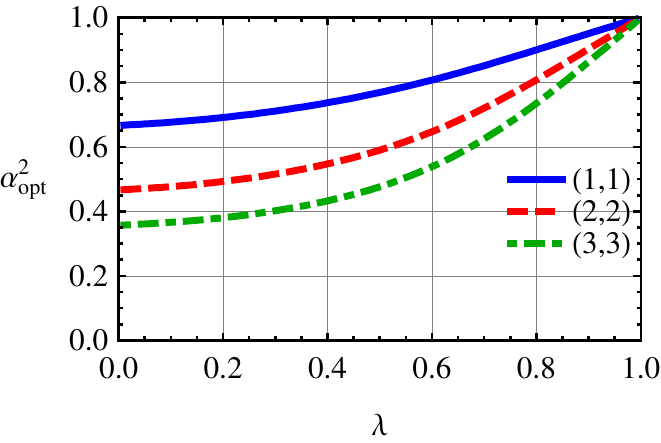}\label{fig:OptAlphaD}}
\subfigure[]{\includegraphics[width=0.35\textwidth]{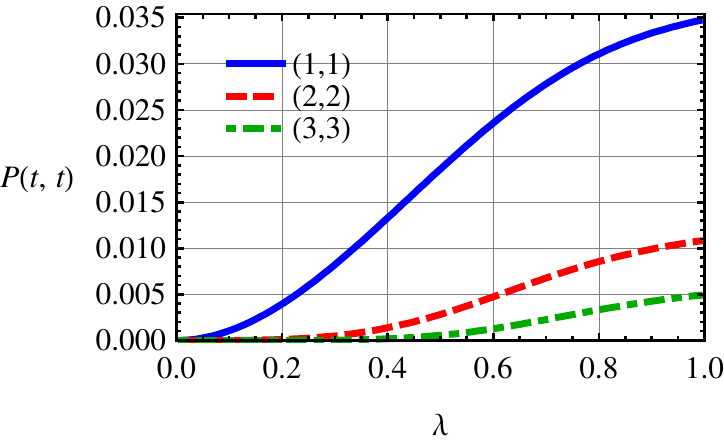}\label{fig:OptProbD}}
\subfigure[]{\includegraphics[width=0.35\textwidth]{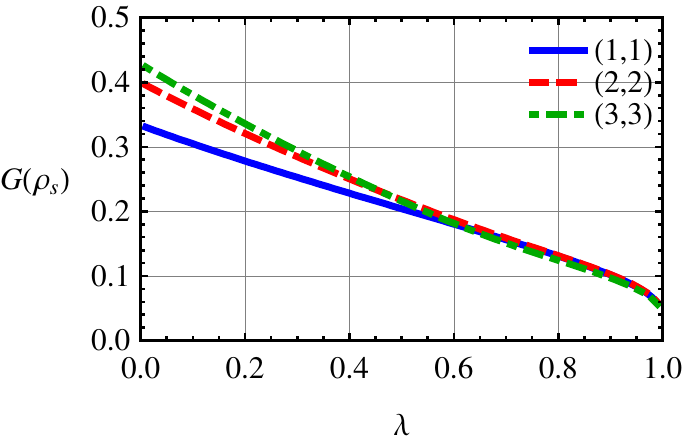}\label{fig:OptGainD}}
\subfigure[]{\includegraphics[width=0.35\textwidth]{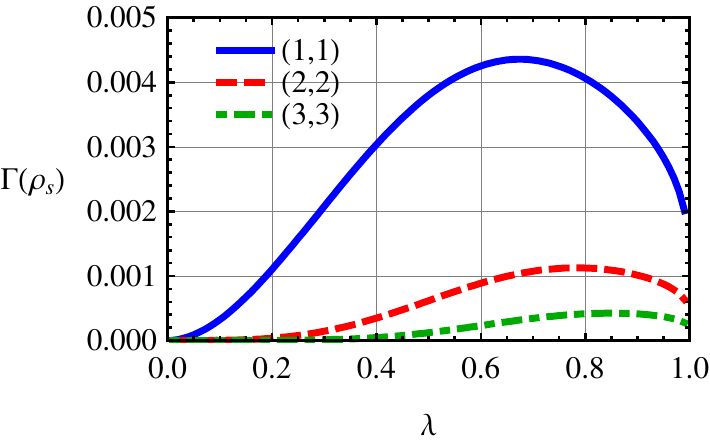}\label{fig:OptRateD}}
\caption{(Color online) Lossless symmetric subtraction with PNRDs. \subref{fig:OptAlphaD}: $\alpha^2$ optimised over entanglement gain rate for a given $\lambda$ for different numbers of subtracted photons $t$.  \subref{fig:OptProbD}, \subref{fig:OptGainD} and~\subref{fig:OptRateD}: the maximum probability, gain  and rate, respectively, achieved with the optimal $\alpha^2$ as a function of $\lambda$, for different values of $t$.}\label{fig:PNRDSymmPerf}
\end{figure}
Perfect symmetric photon subtraction detects an equal number of photons in both modes with 100\% efficiency. This is equivalent to setting $\gamma=\gamma^\prime=0$, whereby $l=l^\prime=0$ and $\alpha=T_2$, the transmissivity of the subtraction beam splitter. Symmetry implies $t=t^\prime$, $\alpha=\alpha^\prime$ and $\beta=\beta^\prime$, implying $K=i+j$, which simplifies Eqn.~(\ref{eqn:Ckij}) to
\be
C_{i,j}^{\left(K\right)}=\frac{\lambda^{K+2t}\alpha^{2K}\beta^{4t}\left(i+t\right)!\left(j+t\right)!}{t!^2 \sqrt{i!j!\left(K-i\right)!\left(K-j\right)!}}\delta_{K,i+j}~,
\ee
where the Kronecker delta function $\delta_{K,i+j}$ means that all off anti-diagonal elements are zero. In this case, each of the $(K+1)\times (K+1)$ blocks $\mathbf{C}^{\left(K\right)}$ are both symmetric, $C_{i,j}^{\left(K\right)}=C_{j,i}^{\left(K\right)}$, and persymmetric $C_{i,j}^{\left(K\right)}=C_{K-i,K-j}^{\left(K\right)},$ which can be exploited~\cite{Zhang_vL11,Cantoni_B76} to compute the log-negativity using
\begin{equation}
E_N\left(\rho\right)=\log_2\left\{\frac{(1-\lambda^2)\sum_{K=0}^\infty\Tr{\mathbf{J}^{\left(K\right)}\mathbf{C}^{\left(K\right)}}}{P\left(t,t\right)}\right\}~,
\end{equation}
where $\mathbf{J}^{\left(K\right)}=[\delta_{i+j,K}]$ is the anti-identity matrix and we include explicitly the normalisation factor in the denominator. The numerator traces over the antidiagonal elements $C_{i,K-i}^{\left(K\right)}$ while the denominator traces over the diagonal elements $C_{i,i}^{\left(K\right)}$. This yields
\begin{align}
&\sum_{K=0}^\infty\Tr{\mathbf{J}^{\left(K\right)}\mathbf{C}^{\left(K\right)}}=  \nonumber\\
&\left(\lambda\beta^2\right)^{2d}\sum_{K=0}^\infty\sum_{i=0}^K\left(\lambda\alpha^2\right)^K\frac{\left(i+t\right)!\left(K-i+t\right)!}{t!^2i!\left(K-i\right)!}\nonumber \\\label{eqn:SymmetricPNRDEnt}
&=\left(\frac{\lambda\beta^2}{1-\lambda\alpha^2}\right)^{2t}\frac{1}{\left(1-\lambda\alpha^2\right)^2}~,
\end{align}
and from Eqn.~(\ref{eqn:TrRho}), the probability simplifies to
\be
\label{eqn:PerfectPNRDTrace}
P\left(t,t\right)=\frac{\left(1-\lambda^2\right)\lambda^{2t}\beta^{4t}}{\left(1-\lambda^2\alpha^4\right)^{t+1}}P_t\left[\frac{1+\left(\lambda\alpha^2\right)^2}{1-\left(\lambda\alpha^2\right)^2}\right]~,
\ee
where $P_n[z]$ is the $n^{\mathrm{th}}$ Legendre polynomial, a special case of the Jacobi polynomial $P_n^{(a,b)}[z]$ found in Eqn.~(\ref{eqn:TrRho}) given by $P_n^{(0,0)}[z]=P_n[z]$  . This is an analytic, closed-form expression of the results in~\cite{Kitagawa_tsc06,Zhang_vL11}, and yields an entanglement of
\begin{equation}
E_N\left(\rho_\mathfrak{s}\right)=\log_2\left\{\left(\frac{1+\lambda\alpha^2}{1-\lambda\alpha^2}\right)^{t+1}\bigg/P_t\left[\frac{1+\left(\lambda\alpha^2\right)^2}{1-\left(\lambda\alpha^2\right)^2}\right]\right\}~.
\end{equation}
Fig.~\ref{fig:SymmSub} depicts the result for varying squeezing parameter $\lambda$ and subtraction coefficient $\alpha^2$. The behaviour of the entanglement gain rate in Fig.~\ref{fig:Rate} shows that there exists an optimum value that provides the best entanglement yield per trial. High gain is less likely, such that the rate is peaked at particular values of $\lambda_\textrm{opt}=0.66$ and $\alpha^2_\textrm{opt}=0.83$. On average, these parameters produce the highest gain in entanglement per trial.

In general, the parameter $\alpha^2$ is freely tunable when performing the protocol, whereas $\lambda$ is restricted by the maximum squeezing available. Therefore it is useful to obtain an expression for the optimal subtraction parameter $\alpha^2_\textrm{opt}$ in terms of the initial squeezing $\lambda$ that maximises the entanglement gain rate $\Gamma\left(\rho_\mathfrak{s}\right)$. These values are obtained from Fig.~\ref{fig:Rate}, and are plotted as a function of $\lambda$ in Fig.~\ref{fig:OptAlphaD}. For $t=1$, the fit to these data is a second-order polynomial of the form
\be
\alpha_\textrm{opt}^2=0.238\left(\lambda-1\right)^2+0.576\left(\lambda-1\right)+1~.
\ee
Thus, given that in the lossless case $\alpha=T_2$, we have the recipe to dial up the most effective subtraction rate to maximise the gain in entanglement per trial.



It has been shown previously that all things being equal, subtracting more photons increases the gain after subtraction~\cite{Zhang_vL11}. However, by optimizing $\alpha^2$ with respect to the rate, this gain is only marginally higher with increasing $d$, as shown in Fig.~\ref{fig:OptGainD}. Indeed, at $\lambda>0.6$, there is no advantage to subtracting more photons. As expected, even with the optimal $\alpha^2$, the probability of entanglement gain decreases significantly with increasing $t$, shown in Fig.~\ref{fig:OptProbD}. These effects combine to yield a lower entanglement gain rate as the number of subtracted photons is increased, as shown in Fig.~\ref{fig:OptRateD}.

\subsection{Lossless asymmetric subtraction}\label{sec:PNRDAsymm}
\begin{figure*}
\centering
\subfigure[]{\includegraphics[width=0.285\textwidth]{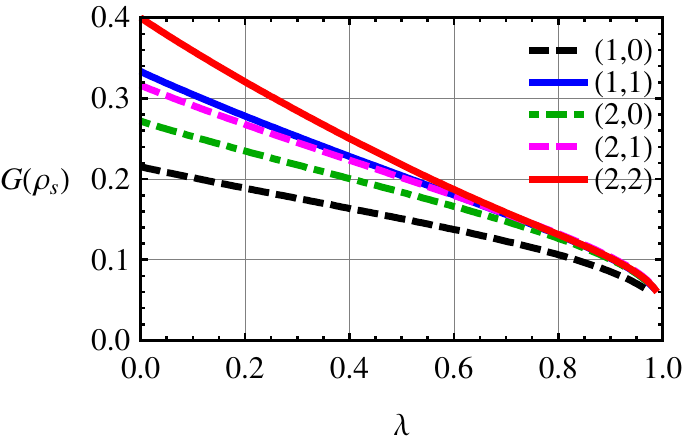}\label{fig:OptGainAsymm}}\qquad
\subfigure[]{\includegraphics[width=0.3\textwidth]{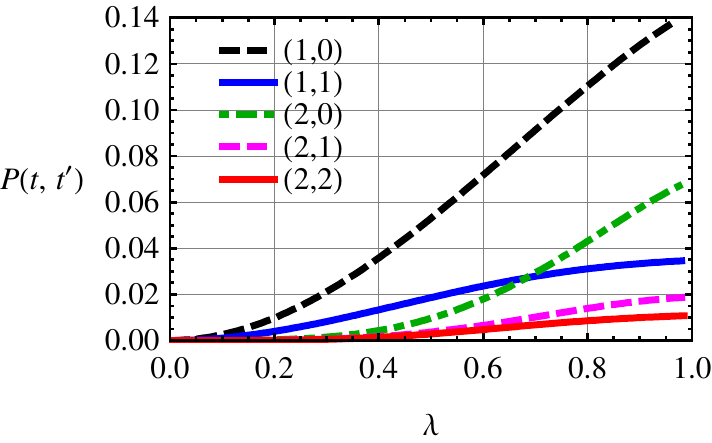}\label{fig:OptProbAsymm}}\qquad
\subfigure[]{\includegraphics[width=0.3\textwidth]{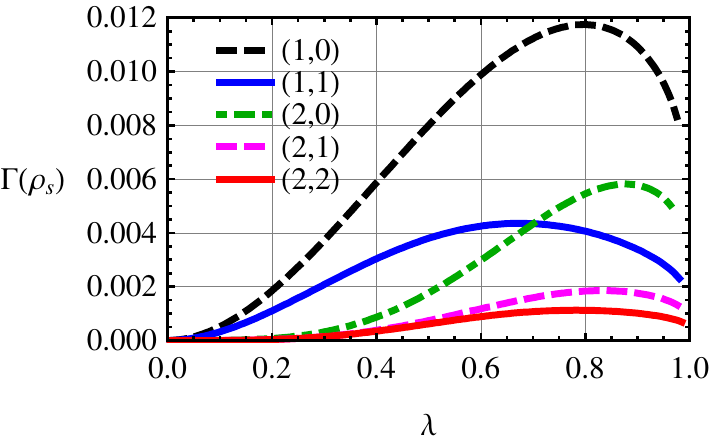}\label{fig:OptRateAsymm}}
\caption{(Color online) Lossless symmetric and asymmetric subtraction with PNRDs. Using $\alpha,\alpha^\prime$ optimised over entanglement gain rate for different combinations of subtracted photons $(t,t^\prime\leq2)$, we calculate~\subref{fig:OptGainAsymm} the achievable gain,~\subref{fig:OptProbAsymm} the success probability and~\subref{fig:OptRateAsymm} entanglement gain rate as a function of $\lambda$.}\label{fig:AsymmParams}
\end{figure*}
We now consider the case of perfect \textit{asymmetric} subtraction, \emph{i.e.}\ detecting different numbers of photons $t\neq t^\prime$ in each subtracted mode.  When counting resources in terms of total photons subtracted, asymmetric subtraction allows us to double our space to include odd photon numbers. This additional degree of freedom lifts the degeneracy in the coefficients, \emph{i.e.}\ the primed quantities may take different values from their unprimed counterparts. This doubles the number of parameters over which the protocol can be optimised.

Starting again from Eqn.~(\ref{eqn:Ckij}), we detect $t,t^\prime$ photons in each mode with unit efficiency. Since we are neglecting the effects of loss in the system,  $\gamma=\gamma^\prime=0$ and as such the only non-zero contribution to the summation over $l$ is the $l=0$ term. Photon number resolution is maintained by setting $t_\textrm{max}=t, t_\textrm{max}^\prime=t^\prime$ in Eqn.~(\ref{eqn:Ckij}) as before, however they need not be equal as in the case above.
In this scheme the relation $\gamma=\gamma^\prime=0$ ensures that all elements of $\mathbf{C}^{\left(K\right)}$, except those that satisfy $i+j=K+t^\prime-t$, are zero. We may thus express the elements of $\mathbf{C}^{\left(K\right)}$ as
\begin{align}
C_{i,j}^{\left(K\right)} =& (1-\lambda^2) \lambda^{i+j+2t}\alpha^{i+j}\alpha^{\prime2K-i-j}\beta^{2t}\beta^{\prime2t^\prime}\\
&\times \frac{(i+t)!(j+t)!}{t!t^\prime! \sqrt{i!j!(K-i)!(K-j)!}} \delta_{i+j,K+t^\prime-t}\nonumber~.
\end{align}
The matrices $\mathbf{C}^{\left(K\right)}$ only have elements along one of their skew diagonals, shifted from the main skew diagonal by $t^\prime-t$. This allows us to define an antidiagonal submatrix $\mathbf{B}^{\left(\bar{K}\right)}$, where $\mathbf{B}^{\left(\bar{K}\right)}$ is a $\left(\bar{K}+1\right) \times \left(\bar{K}+1\right)$ matrix and where $\bar{K} = K-t^\prime+t$, which contains the elements on its main skew diagonal. Without loss of generality, we are able to choose $t \leq t^\prime$, whereby the elements of $\mathbf{B}^{\left(\bar{K}\right)}$ are $B_{i,j}^{\left(\bar{K}\right)} = C_{i+t^\prime-t,j+t^\prime-t}^{\left(K\right)}$. 
This is both symmetric and persymmetric, so the above approach can again be used in computing the negativity. When evaluating $\Tr{\mathbf{B}^{\left(\bar{K}\right)} \mathbf{J}^{\left(\bar{K}\right)}}$ the elements of interest are $B_{i,\bar{K}-i}^{\left(\bar{K}\right)} = C_{i+t^\prime-t,K-i}^{\left(K\right)}$, where
\begin{align}\label{eqn:CKSubmatrix}
B_{i,\bar{K}-i}^{\left(\bar{K}\right)}=&(1-\lambda^2) \lambda^{\bar{K}+2t^\prime}\alpha^{\bar{K}+2(t^\prime-t)}\alpha^{\prime\bar{K}} \beta^{2t}\beta^{\prime2t^\prime}\\
&\times \frac{(i+t^\prime)!(\bar{K}-i+t^\prime)!}{t!t^\prime! \sqrt{(i+t^\prime-t)!(\bar{K}-i+t^\prime-t)!(\bar{K}-i)!i!}}~. \nonumber
\end{align}
Eqns.~(\ref{eqn:CKSubmatrix},\ref{eqn:TrRho}) together lead to the relations
\begin{align}\nonumber
P\left(t,t^\prime\right)=& \frac{\left(1-\lambda^2\right)\lambda^{2t^\prime} \beta^{\prime2t^\prime} \beta^{2t} \alpha^{2\left(t^\prime-t\right)} }{ \left(1- \alpha^2 \alpha^{\prime2} \lambda^2 \right)^{t^\prime+1} } \\\label{eqn:BkJk}
&\times P_t^{\left(t^\prime-t,0 \right)} \left[ \frac{1+\left( \alpha \alpha^\prime \lambda \right)^2 }{1-\left( \alpha \alpha^\prime \lambda \right)^2 } \right]~,
\end{align}
and
\begin{align}
\sum_{\bar{K}=0}^{\infty} \Tr{\mathbf{B^{\left(\bar{K}\right)} \mathbf{J}^{\left(\bar{K}\right)}}}=&(1-\lambda^2) ( \lambda \beta^\prime)^{2t^\prime} \beta^{2t} \alpha^{2\left(t^\prime-t\right)} \nonumber \\
&\times \left( \sum_{i=0}^{ \infty} (\lambda \alpha \alpha^\prime)^i \frac{\left(i+t^\prime\right)!}{\sqrt{i!\left(i+t^\prime-t\right)!t!t^\prime!}} \right)^2~.
\end{align}
Unfortunately, no analytical expression could be found for the sum of series in brackets. However, by considering the ratios between successive terms in the series, it is clear that it converges over the relevant range $0 < \lambda \alpha \alpha^\prime < 1$. This gives an exact, but not closed, form for the log-negativity of the state
\begin{align}\nonumber
E_N(\rho_\mathfrak{s})=& \log_2 \left\{ \frac{ \left(1- \alpha^2\alpha^{\prime2}\lambda^2\right)^{t^\prime+1}}{P_t^{\left(t^\prime-t,0 \right)} \left[ \frac{1+\alpha^2 \alpha^{\prime2} \lambda^2 }{1-\alpha^2\alpha^{\prime2}\lambda^2 } \right]} \right\} \\
&+2 \log_2 \left\{ \sum_{i=0}^{ \infty} \left(\lambda \alpha \alpha^\prime\right)^i \frac{(i+t^\prime)!}{\sqrt{i!\left(i+t^\prime-t\right)!t!t^\prime!}} \right\}~.
\end{align}

\subsubsection{Comparing symmetric and asymmetric subtraction}
As with the symmetric case earlier, we can directly compare the entanglement gain, probability and rate for asymmetric subtraction. For illustration purposes, we compare different ways of subtracting up to two photons from each mode. The results are shown in Fig.~\ref{fig:AsymmParams}.

It is clear from Fig.~\ref{fig:OptGainAsymm} that the symmetric cases, $(t,t\prime)=(1,1)$ and $(2,2)$, produce more gain than their asymmetric counterparts for a fixed number of subtracted photons. From Fig.~\ref{fig:OptProbAsymm}, there exists a regime of $\lambda \gtrsim 0.7$ for which the probability of subtracting (2,0) photons is greater than the (1,1) case, therefore the gain rate, shown in Fig.~\ref{fig:OptRateAsymm} is correspondingly higher in this regime. This is because the (2,0) subtraction event is more likely than the (1,1) for high values of $\lambda$. However, it is clear that in general the asymmetric (1,0) case produces the most entanglement gain per trial, once again due to the high likelihood of subtracting 1 photon as opposed to 2 photons in any combination.

\subsection{Losses and imperfect detection}
\label{sec:losses}

\begin{figure*}
\centering
\subfigure[]{\includegraphics[width=0.37\textwidth]{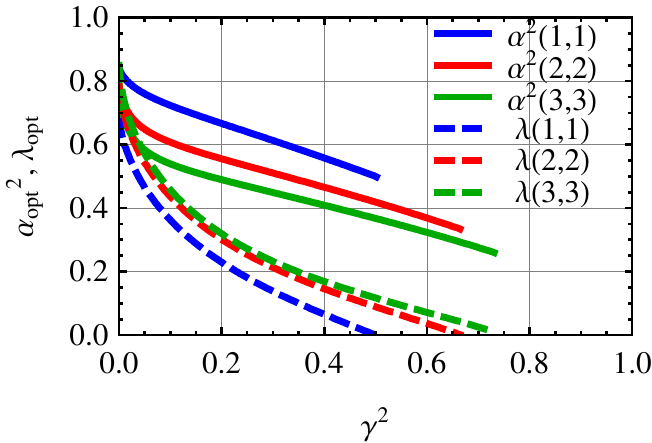}\label{fig:LossOptParam}}\qquad
\subfigure[]{\includegraphics[width=0.4\textwidth]{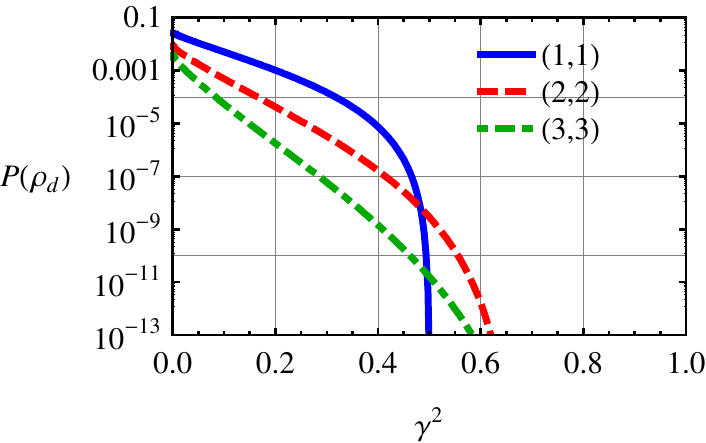}\label{fig:LossRate}}\\
\subfigure[]{\includegraphics[width=0.4\textwidth]{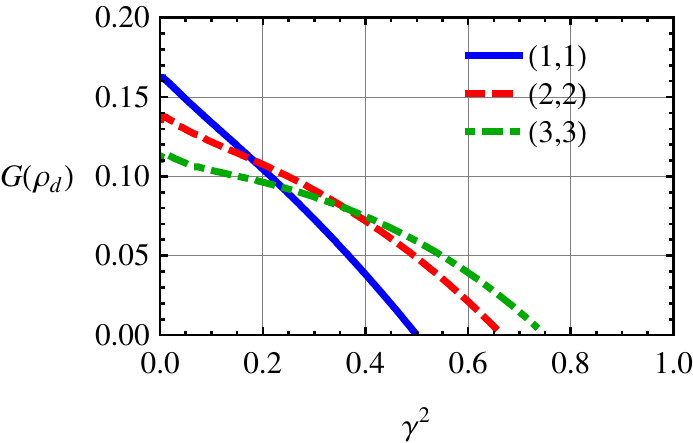}\label{fig:LossGain}}\qquad
\subfigure[]{\includegraphics[width=0.4\textwidth]{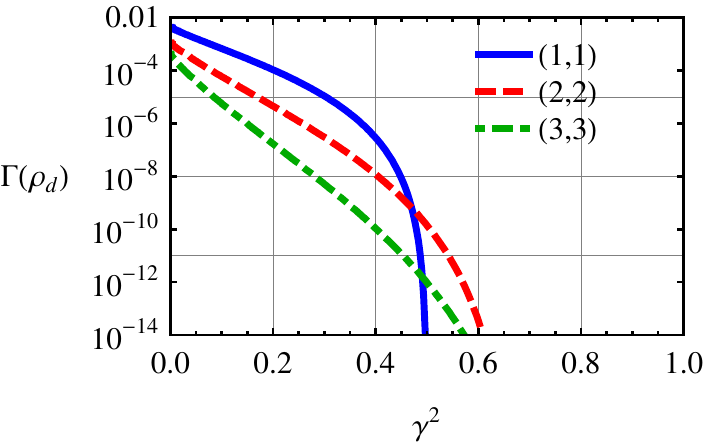}\label{fig:LossEff}}
\caption{(Color online) Lossy symmetric subtraction with PNRDs. \subref{fig:LossOptParam}: squeezing $\lambda$ and subtraction $\alpha^2$ parameters, optimised to produce the highest entanglement gain rate at a given loss; and the resulting \subref{fig:LossGain}~gain, \subref{fig:LossRate}~probability and \subref{fig:LossEff}~rate resulting from these parameters.}\label{fig:PNRDSymmLoss}
\end{figure*}

In any realistic scenario, where the process of entanglement distillation and entanglement enhancement will be most essential, the detectors are imperfect and evolving quantum states suffer losses. Losses are accounted for by setting  $\gamma,\gamma^\prime > 0,$ and the summation over $l$ (which denotes the number of photons lost) includes contributions from $l\geq 1$ which are non-zero. Losses are modelled by beam splitters before, during and after subtraction, as shown in Fig.~\ref{fig:Subtr_setup}. 
The effect of loss, for the symmetric cases where one (blue), two (red) and three (green) photons are detected from each mode, is shown in Fig.~\ref{fig:PNRDSymmLoss}. 
As the combined loss $\gamma^2$ increase, optimal initial squeezing $\lambda$ for which entanglement can be enhanced by local photon subtraction decreases. Furthermore, detecting higher numbers of photons is more loss-tolerant; since a smaller $\alpha^2$ is required to subtract more photons, the restriction on $\gamma^2\leq0.5$ is reduced. Indeed, we can define $\gamma^2_\textrm{max}$ as the maximum loss for which entanglement still increases on subtraction. The dependence of the maximum losses $\gamma_\textrm{max} = 1/t+1$ comes from the gain as a function of loss, setting to zero and solving for $t$ the difference of Eqns.~(\ref{eqn:EntPNRDSymmLoss}) and~(\ref{eqn:EntInit}).

The constituents of $\gamma,\gamma^\prime$ are determined by Eqn.~(\ref{eqn:Gamma}), where we define efficiencies before, during and after detection as $T_1^2,~T_3^2,$ and $T_4^2$ respectively. From the expressions of $\gamma,\gamma^\prime$, it is clear that each of these efficiencies contribute differently to losses. 
The effect of detector efficiency, parameterised by $T_3$, is almost negligible. This can be interpreted by considering the subtraction detector as something of a post-selector: it may not often click, but when it does one can be fairly certain that entanglement has been increased. Losses after subtraction may be mitigated in a similar way, by using a loss-tolerant entanglement detection strategy, such as post-selection or a loss-tolerant entanglement witness~\cite{Eisert07wit}. Loss before subtraction, which may be modelled as a mixing of the state due to non unit channel transmission, is unavoidable and its contribution is significant.

\subsubsection{Entanglement gain and rate under loss}
To study the enhancement of entanglement in the presence of losses and imperfection, we again start from Eqn.~(\ref{eqn:Ckij}), which in the symmetric detection PNRD case for $t=t^\prime$, is given by
\begin{align}
C_{i,j}^{\left(K\right)}=&\sum_{l=l_0}^\infty\lambda^{i+j+2\left(t+l\right)}\alpha^{2K}\beta^{4t}\gamma^{2\left(i+j-K+2l\right)}\\
&\times\frac{\left(i+l+t\right)!\left(j+l+t\right)!}{l!\left(l+i+j-K\right)!t!^2\sqrt{i!j!\left(K-j\right)!\left(K-i\right)!}}\nonumber~,
\end{align}
where $l_0=\max\left(0,K-i-j\right)$. Since this matrix is symmetric and persymmetric, we can calculate the entanglement following~\cite{Zhang_vL11}. Defining $x=1-\alpha^2\lambda$ and $y=\left(\alpha^2+\gamma^2\right)\lambda$ allows us to write
\begin{align}
\Tr{\textbf{J}^{\left(\textrm{K}\right)}\textbf{C}^{\left(\textrm{K}\right)}}=&\frac{ \left(1-\lambda^2\right) \beta^{4t} \lambda^{2t} }{ \left( x^2 - \gamma^4 \lambda^2 \right)^{t+1} }P_t \left[ \frac{x^2+\gamma^4\lambda^2}{x^2-\gamma^4 \lambda^2} \right]~,\nonumber
\end{align}
and the success probability is given by
\be\label{eqn:ProbPNRDSymmLoss}
P\left(t,t\right)=\frac{ \left(1-\lambda^2 \right) \beta^{4t} \lambda^{2t}}{ \left[1-y^2\lambda^2 \right]^{t+1} }
P_t \left[ \frac{1+y^2 }{1-y^2 } \right]~.
\ee
 This yields a log-negativity of
\be\label{eqn:EntPNRDSymmLoss}
E_N(\rho)=\log_2 \left[ \left( \frac{1-y^2}{x^2 - \gamma^4\lambda^2 } \right)^{t+1} \frac{P_t \left[ \frac{x^2+\gamma^4\lambda^2}{x^2-\gamma^4\lambda^2} \right]}{P_t \left[ \frac{1+y^2 }{1-y^2 } \right]} \right]~.
\ee
These results reduce to the lossless symmetric detection case given by Eqns.~(\ref{eqn:SymmetricPNRDEnt}, \ref{eqn:PerfectPNRDTrace}) when $\gamma=\gamma^\prime=0$.

\subsubsection{Optimising parameters under loss}\label{sec:PNRDSymmLoss}
Subtraction is a general strategy for increasing entanglement. However, it is interesting to investigate which two-mode squeezed states, parameterised by $\lambda$, are most improved by this scheme. Furthermore, using Eqns.~(\ref{eqn:ProbPNRDSymmLoss}) and~(\ref{eqn:EntPNRDSymmLoss}), we can incorporate loss as an additional input parameter. For a given implementation, the experimentalist has free choice of the reflectivity of the subtraction beam splitter (proportional to $\alpha$). Therefore a logical question to ask would be: given losses of a certain level, what is the optimum range of $\lambda$ for which subtraction yields the highest entanglement gain rate, and to what value should the subtraction beam splitter reflectivity be set in order to achieve this?


Figure~\ref{fig:LossOptParam} shows the pairs of $\lambda$ and $\alpha^2$ values required to produce the highest entanglement gain rate in the presence of losses. Finding an empirical fit to these data one obtains the following formulae for setting the optimal $\lambda$ and $\alpha^2$:
\begin{align}
\alpha_\textrm{opt}\left(\gamma^2\right)&=e^{-38.1\left(\gamma^2+0.1\right)}-0.6\gamma^2+0.8~,\\
\lambda_\textrm{opt}\left(\gamma^2\right)&=e^{-107.1\left(\gamma^2+0.1\right)^2}+e^{-2.8\left(\gamma^2+0.1\right)}-0.2~.
\end{align}
The gain and rate resulting from these parameters are plotted as functions of the loss parameter $\gamma$ in Figs.~\ref{fig:LossGain} and~\ref{fig:LossEff} respectively. One can immediately see that there is an upper threshold on loss, above which entanglement cannot be increased.


\subsubsection{Symmetric subtraction, asymmetric loss}
\begin{figure*}
\centering
\subfigure[]{\includegraphics[width=0.32\textwidth]{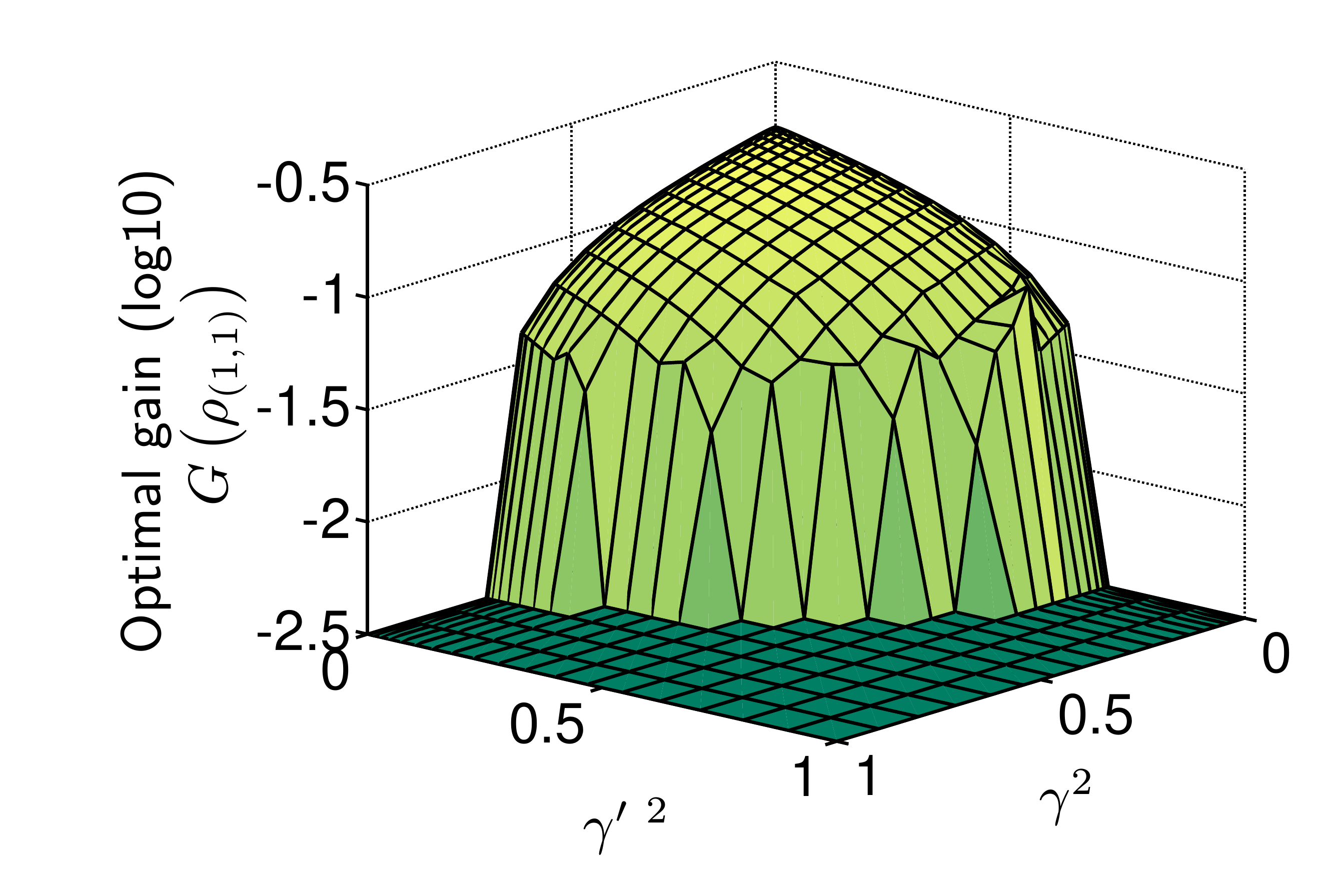}\label{fig:SymmLossGain}}
\subfigure[]{\includegraphics[width=0.32\textwidth]{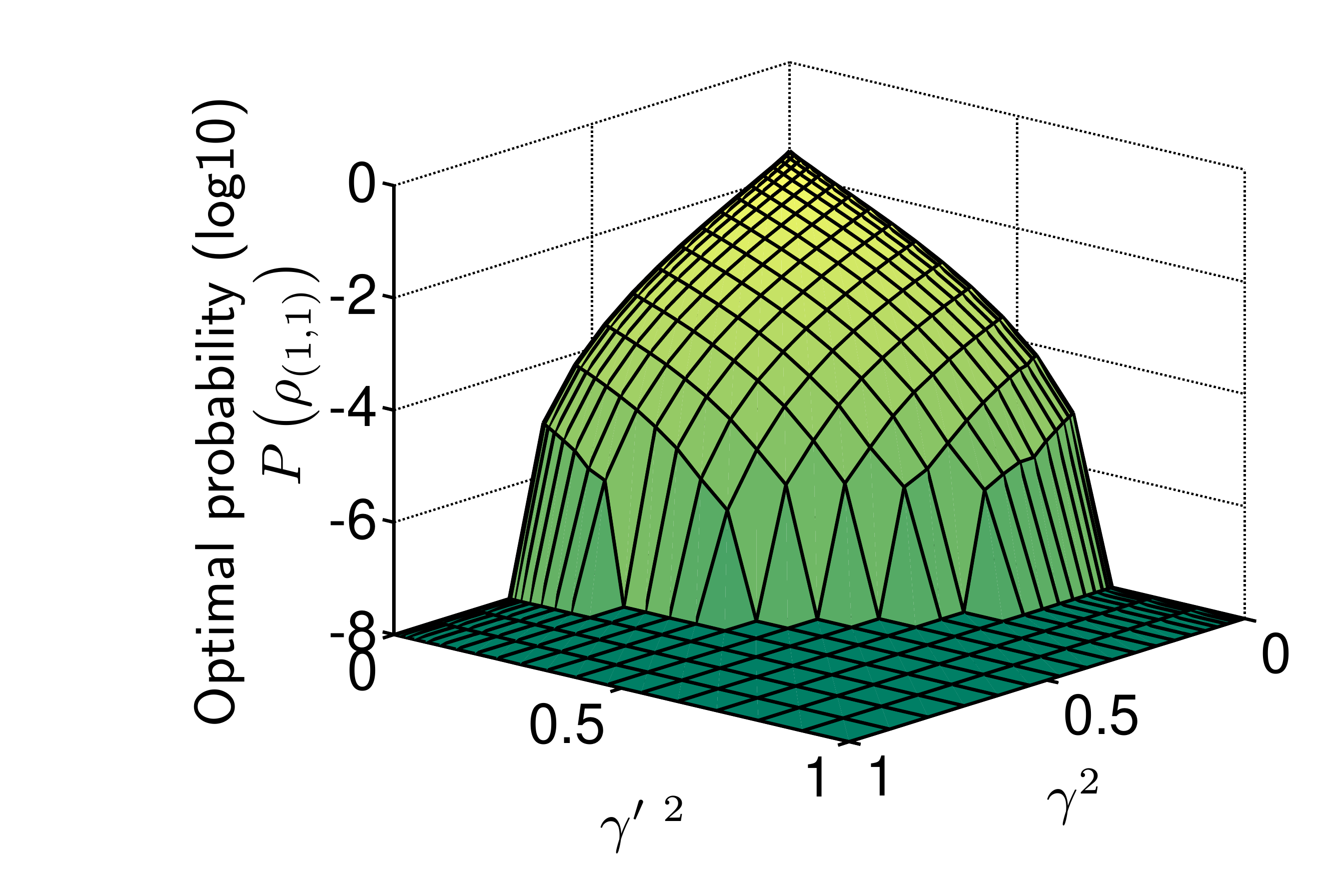}\label{fig:SymmLossTrace}}
\subfigure[]{\includegraphics[width=0.32\textwidth]{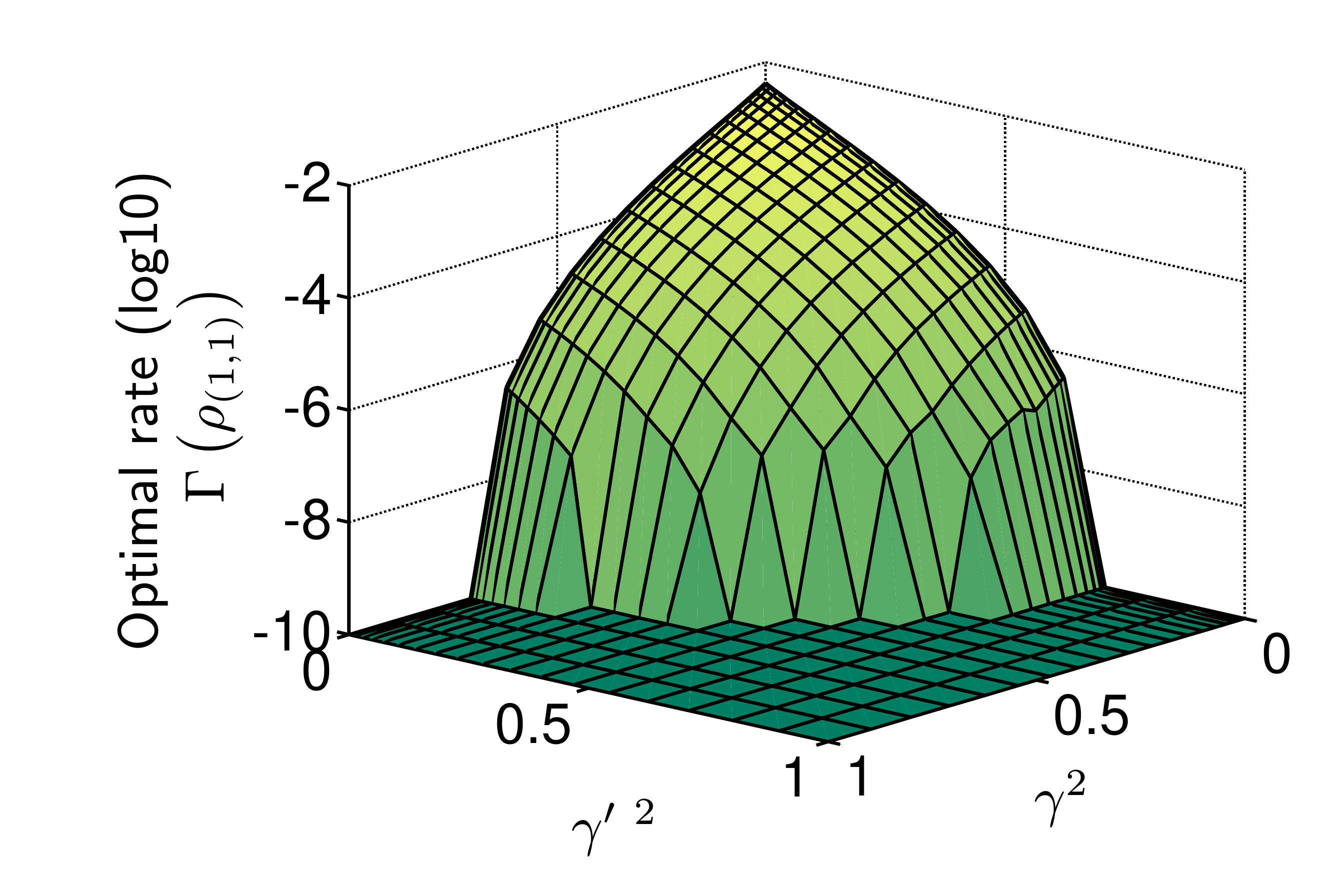}\label{fig:SymmLossDE}}
\caption{(Color online) Lossy symmetric subtraction with PNRDs. The \subref{fig:SymmLossGain}~gain, \subref{fig:SymmLossTrace}~probability and \subref{fig:SymmLossDE}~rate of entanglement increase as a function of loss $\gamma^2,\gamma^{\prime2}$, given one photon subtracted from each mode.}\label{fig:PNRDSymm}
\end{figure*}
By setting $t^\prime=t$ in Eqn.~(\ref{eqn:TrRho}), one may calculate the probability of symmetric subtraction allowing for different parameters $\alpha,\beta,\gamma$ on each mode. However, lifting the degeneracy of the parameters across the different modes means the submatrices $\mathbf{C}^{\left(\mathbf{K}\right)}$ are no longer centrosymmetric, and therefore the entanglement is not analytically tractable. Instead we calculate numerically the entanglement from Eqn.~(\ref{eqn:Ckij}).
Fig.~\ref{fig:PNRDSymm} shows the gain, probability and rate for non-degenerate losses when subtracting a single photon from each mode. The results in Fig.~\ref{fig:PNRDSymmLoss} are simply line-outs for $\gamma=\gamma^{\prime}$.

\subsection{Losses in asymmetric subtraction}\label{sec:PNRDAsymmLoss}

\begin{figure*}
\centering
\subfigure[]{\includegraphics[width=0.45\textwidth]{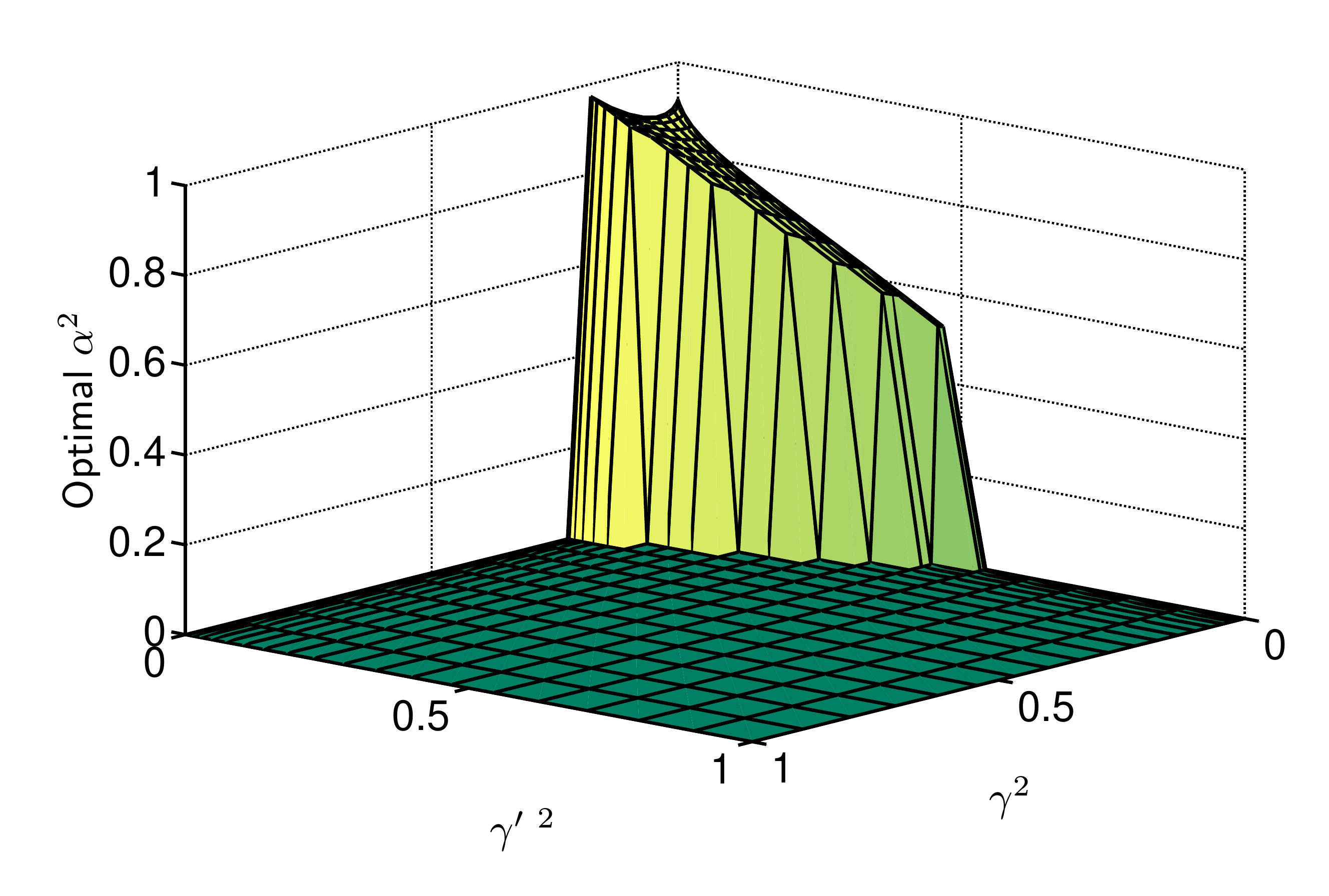}\label{fig:AsymmLossA1Opt}}\qquad
\subfigure[]{\includegraphics[width=0.45\textwidth]{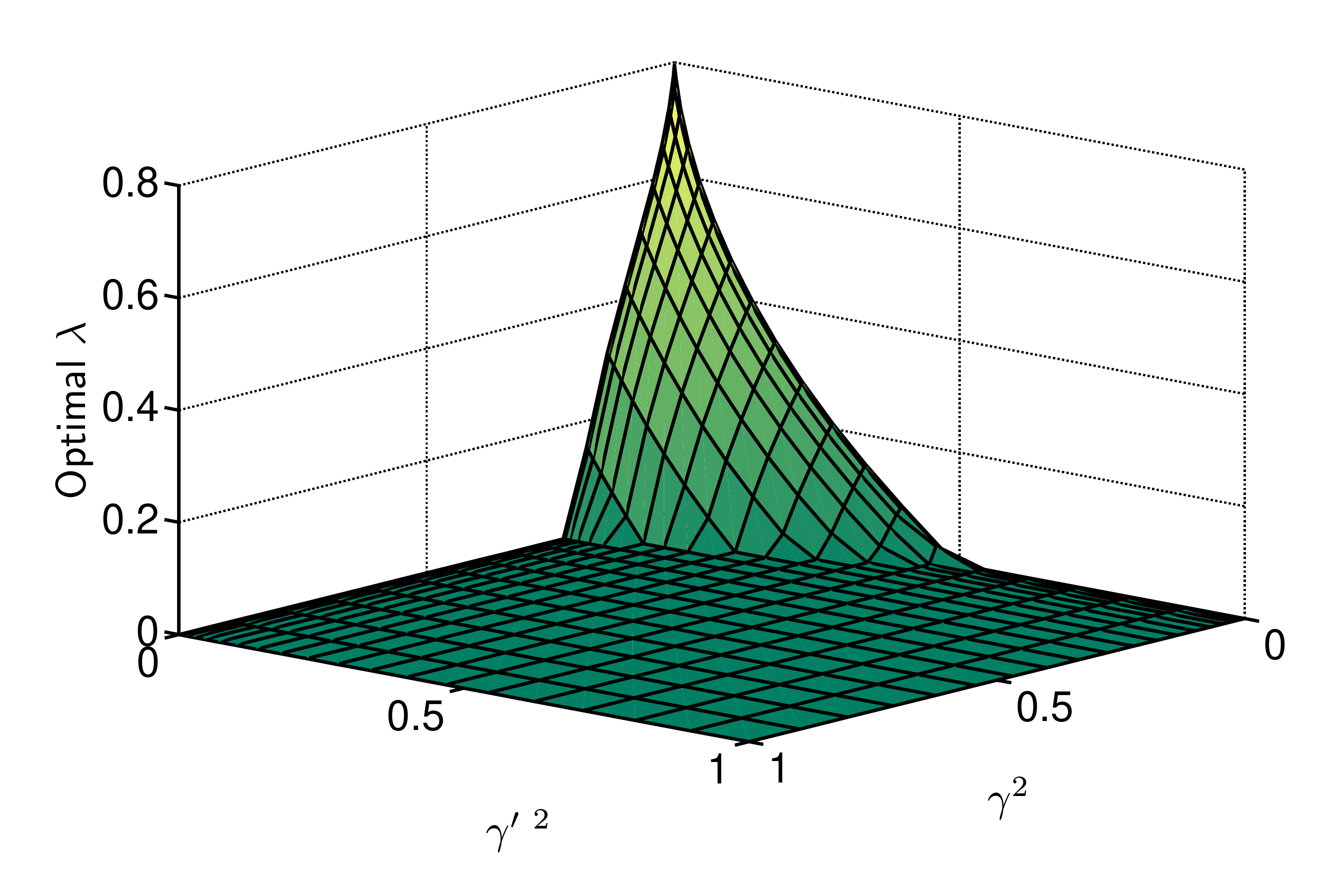}\label{fig:AsymmLossLambdaOpt}}
\caption{(Color online) Lossy asymmetric subtraction with PNRDs. \subref{fig:AsymmLossA1Opt}:~optimised subtraction parameter $\alpha^2_\textrm{opt}$ and \subref{fig:AsymmLossLambdaOpt}:~squeezing parameter $\lambda_\textrm{opt}$ under loss given asymmetric (1,0) subtraction.}\label{fig:PNRDAsymmLoss}
\end{figure*}

When subtracting photons asymmetrically, the effect of loss on each mode may not be equivalent. As the simplest example of asymmetric subtraction, we consider the $(t,t^\prime)=(1,0)$ case, whereby one photon is subtracted from one mode, and the other mode is left unchanged. As shown in Fig.~\ref{fig:OptRateAsymm}, this case produces the highest entanglement gain rate for all $\lambda$. We calculate numerically the optimal parameters $\alpha^2_\textrm{opt},~\lambda_\textrm{opt}$ that give the largest rate of gain, as shown in Fig.~\ref{fig:PNRDAsymmLoss}. Note that, since we do not subtract from the second mode, the optimal $\alpha^{\prime2}$ is, trivially, $1-\gamma^{\prime2}$ for all $\gamma^2$.

\begin{figure*}
\centering
\subfigure[]{\includegraphics[width=0.32\textwidth]{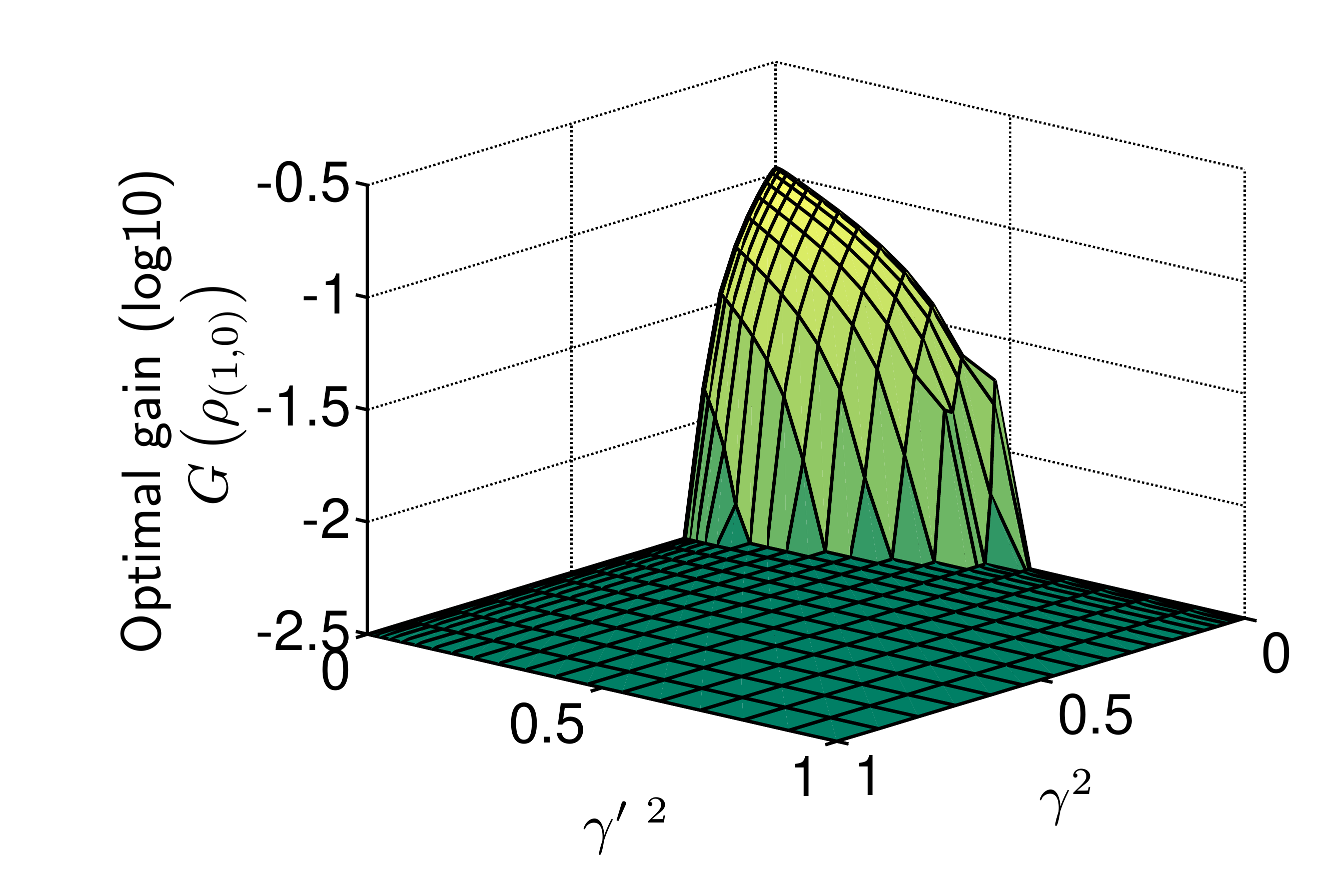}\label{fig:AsymmLossGain}}
\subfigure[]{\includegraphics[width=0.32\textwidth]{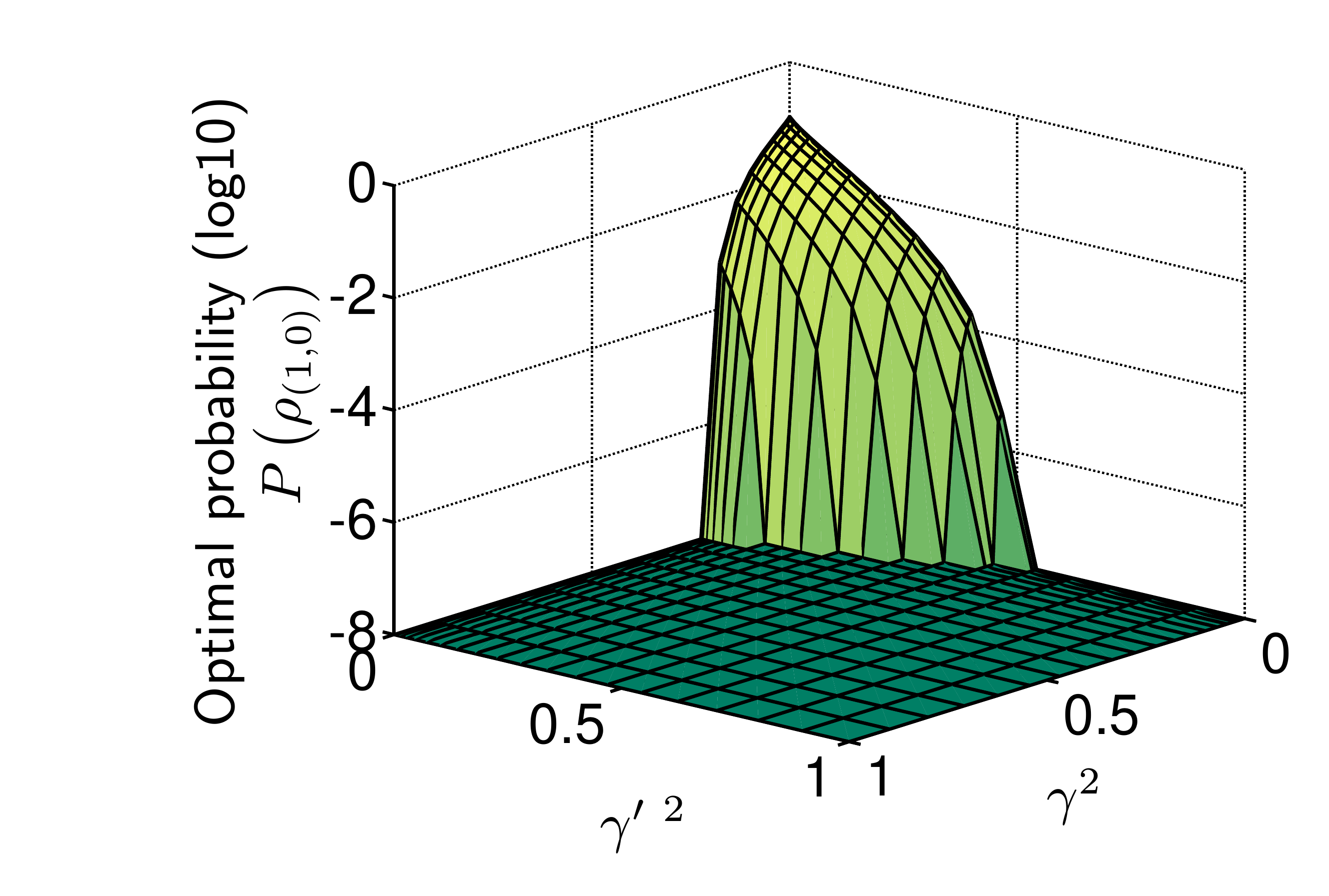}\label{fig:AsymmLossTrace}}
\subfigure[]{\includegraphics[width=0.32\textwidth]{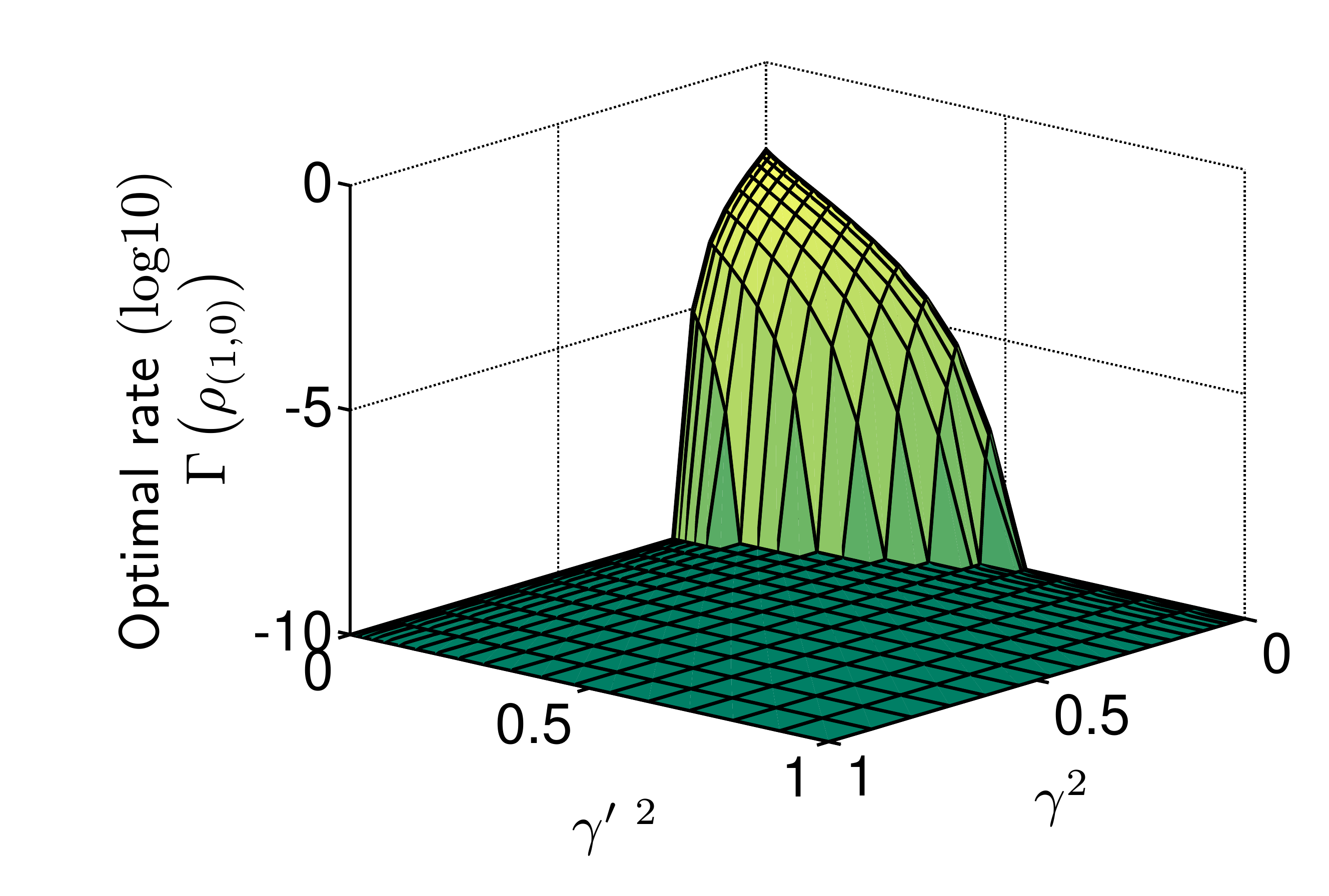}\label{fig:AsymmLossDE}}
\caption{(Color online) Lossy asymmetric subtraction using PNRDs.  \subref{fig:AsymmLossGain}~gain, \subref{fig:AsymmLossTrace}~probability and \subref{fig:AsymmLossDE}~rate of entanglement gain as a function of the losses $\gamma^2,\gamma^{\prime2}$ on modes one and two respectively. The probability of successful subtraction decreases very quickly with loss, and the positive gain region is bounded asymmetrically, being roughly twice as sensitive to losses on the unsubtracted mode.}\label{fig:PNRDAsymm}
\end{figure*}

The resulting gain, probability and rate are shown in Figs.~\ref{fig:AsymmLossGain}, \subref{fig:AsymmLossTrace} and~\subref{fig:AsymmLossDE} respectively. As is to be expected, the effect of loss is not symmetric in this case. The gain region is bounded more sharply by losses in the unsubtracted mode, whereas the losses in the subtracted mode follow the scaling of the symmetric case, Fig.~\ref{fig:LossOptParam}. 


\section{Subtraction with threshold detectors}\label{sec:APD}
\begin{figure*}
\centering
\subfigure[]{\includegraphics[width=0.3\textwidth]{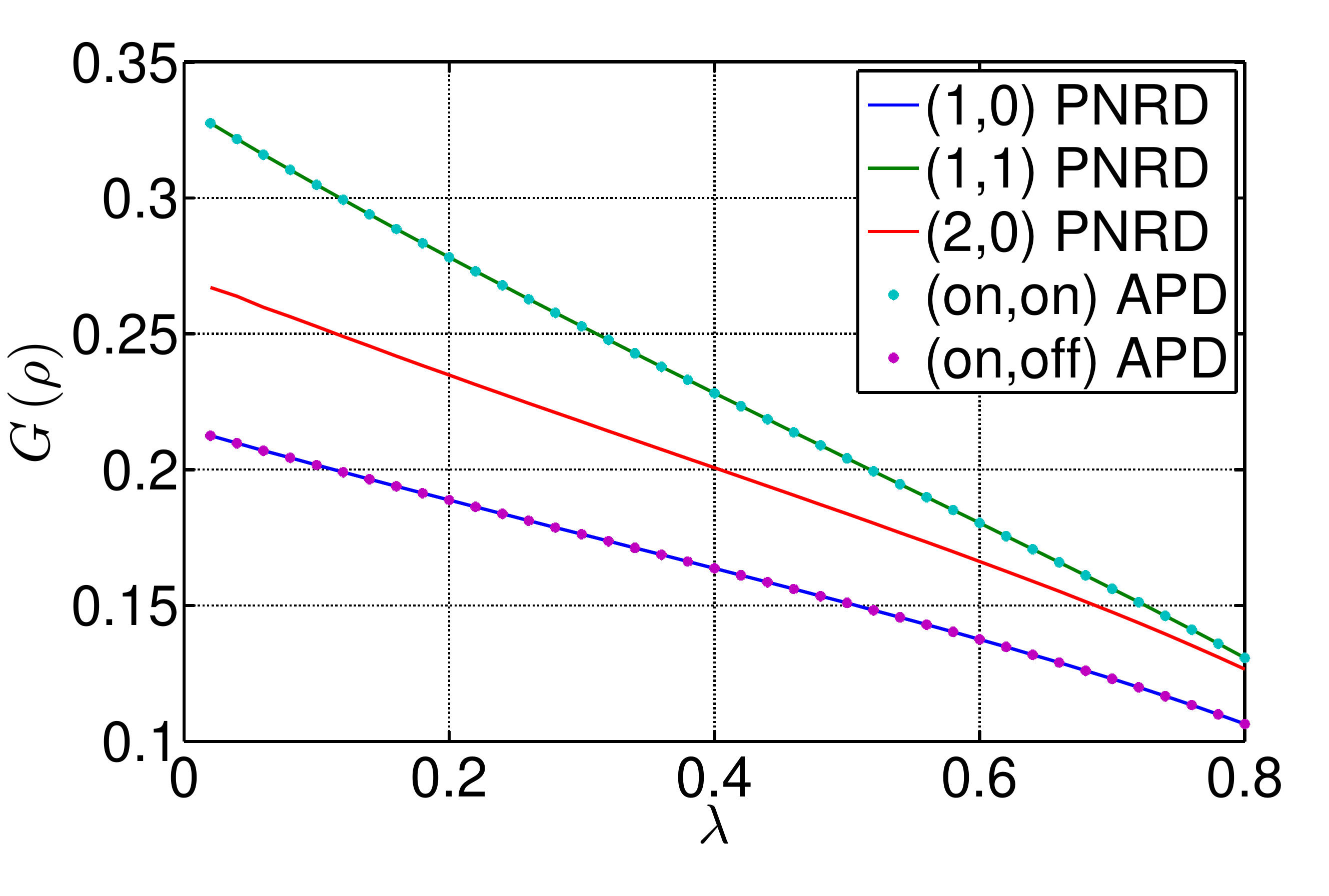}\label{fig:APD_PNRD_gain}}\qquad
\subfigure[]{\includegraphics[width=0.3\textwidth]{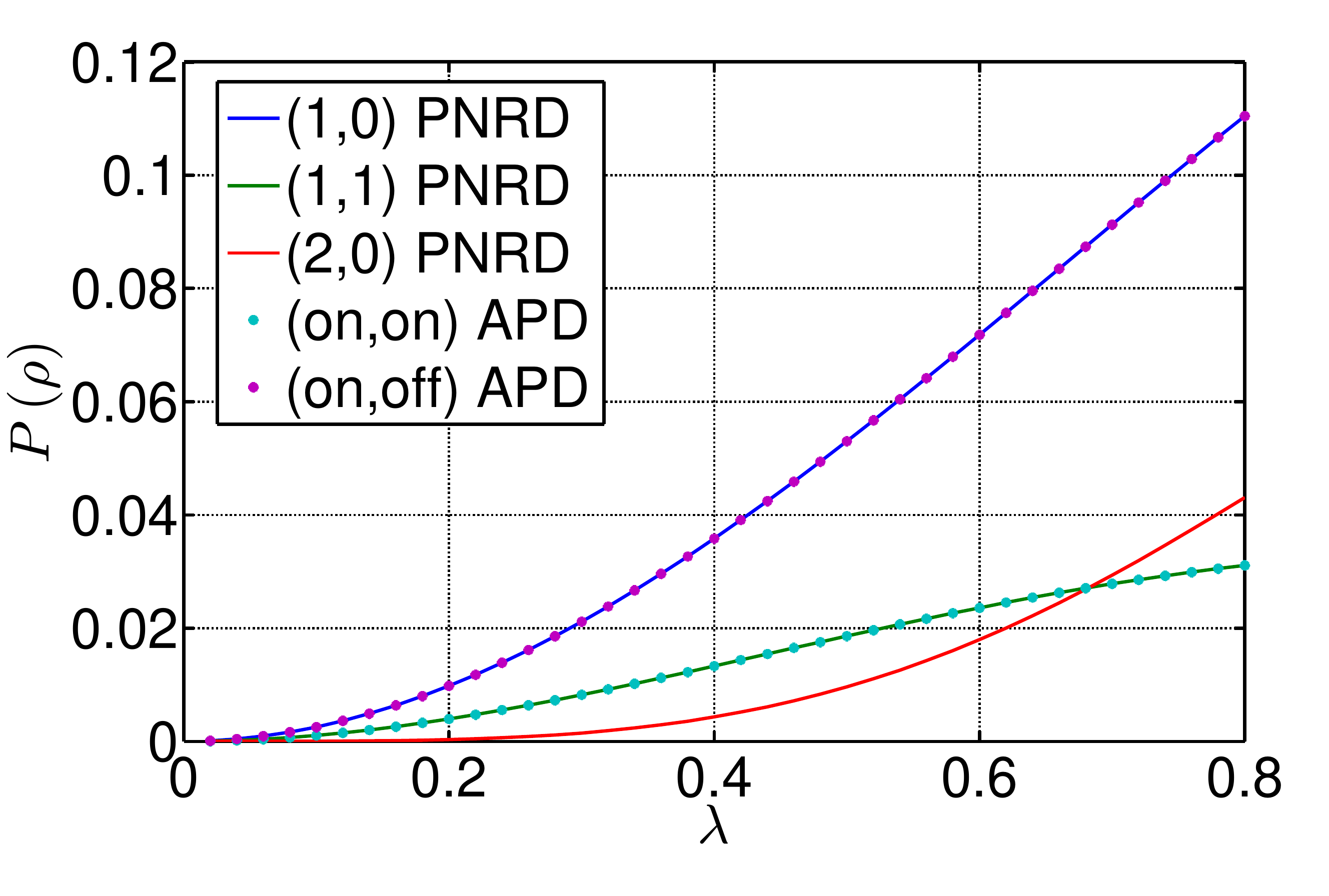}\label{fig:APD_PNRD_prob}}\qquad
\subfigure[]{\includegraphics[width=0.3\textwidth]{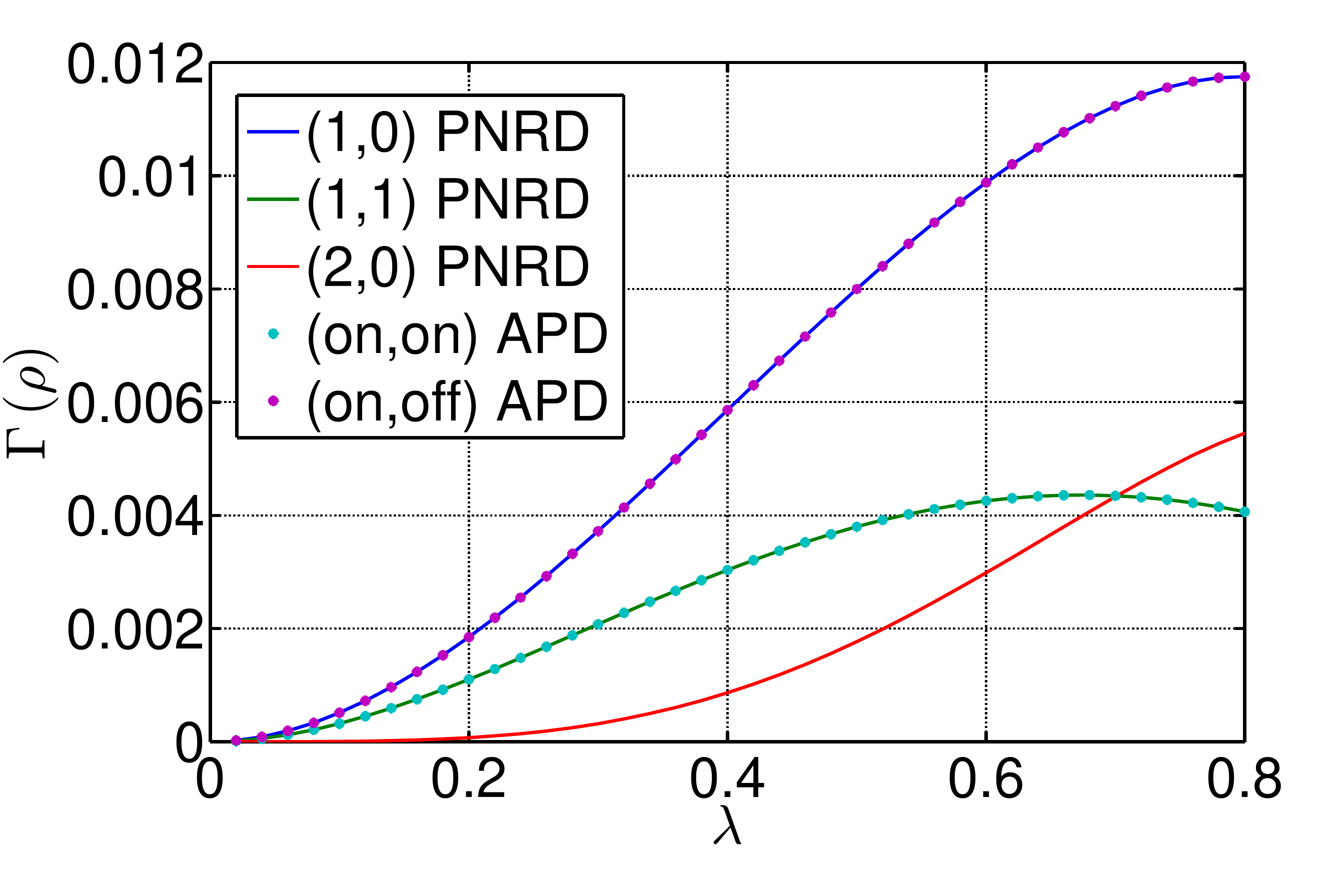}\label{fig:APD_PNRD_rate}}
\caption{(Color online) Comparing lossless subtraction strategies between APDs and PNRDs. \subref{fig:APD_PNRD_gain}~subtraction gain, \subref{fig:APD_PNRD_prob}~probability and \subref{fig:APD_PNRD_rate}~rate for lossless PNRDs and APDs in various subtraction strategies. The results for APDs and PNRDs for simialr strategies, up to $\lambda=0.8$, are identical.}\label{fig:APD_PNRD_comparison}
\end{figure*}
We now proceed to study the enhancement of entanglement that can be effected by subtracting photons locally using threshold detectors. We consider avalanche photodiodes (APDs), detectors that click when at least one photon is incident. Starting from Eqn.~(\ref{eqn:Ckij}), the effect of such detectors are described by the summations over $d$ from $t=t_\textrm{max}=0$ for ``off'' or from $t=1$ to $t_\textrm{max}=\infty$ for ``on'', and similarly for $t^\prime$. This analysis can be extended to higher photon number threshold detectors by changing the ranges corresponding to the ``on'' and ``off'' detection. A physical example of such a detector would be a multiplexed array of $n$ APDs with a detection event triggered off $n$-fold coincidences.



\subsection{Probability of photon subtraction using APDs}
The four measurement outcomes to consider when using APDs are as follows: both detectors clicking (on,on), neither clicking (off,off) or just one  clicking (on,off) or (off,on). The probabilities corresponding to these events must sum to unity.

The (off,off) case is identical to the case where both PNRDs register no photons. Defining $y=\left(\alpha^2+\gamma^2\right)\lambda,~y^\prime=\left(\alpha^{\prime 2}+\gamma^{\prime 2}\right)\lambda$, the probability in this case is
\be
P(\textrm{off,off})= \frac{1-\lambda ^2}{1-yy^\prime}~.
\ee
The (off,on) case is also analytically tractable when incorporating losses. From Eqn.~(\ref{eqn:TrRho}), $P\left(\textrm{off,on}\right)$ is given by
\begin{align}\nonumber
P\left(\textrm{off,on}\right)&=\sum_{t^\prime=1}^\infty P\left(0,t^\prime\right)\\
&=\frac{\left(1-\lambda^2\right)y\left(\lambda-y^\prime\right)}{\left(1-y\lambda\right)\left(1-yy^\prime\right)}~.
\end{align}
Note that $P\left(\textrm{on,off}\right)$ is identical to the result above under exchange of primed and unprimed parameters.
The probability of both detectors registering photons $P\left(\textrm{on,on}\right)$ can be tackled by using the relation $\sum_{p,q=(\mathrm{on,off})}P\left(p,q\right)=1$, yielding
\begin{align}\label{eqn:TrRhoOnOn}
P\left(\textrm{on,on}\right) =&\frac{\left(y-\lambda\right)\left(y^\prime-\lambda\right)\left(1-y y^\prime\lambda^2\right)}{\left(1-y\lambda\right)\left(1-y^\prime\lambda\right)\left(1-yy^\prime\right)}~.
\end{align}

\subsection{Lossless subtraction}\label{sec:APDPerf}

The idealised case of lossless symmetric subtraction using APDs with unit efficiency is analytically tractable~\cite{Zhang_vL11}. However, lifting the degeneracy of the parameters means the submatrices $\mathbf{C}_{K}$ are no longer centrosymmetric, rendering intractable the analytical method used previously. All subsequent results are therefore numerical.

In Ref.~\cite{Zhang_vL11}, it was shown that for squeezing parameters $\lambda<0.95$,  the results are essentially equivalent to those with PNRDs. The same is true for perfect asymmetric subtraction: the case where (1,0) photons are simultaneously detected on each mode respectively, is, within the limits of our numerical analysis, identical to the case where (on,off) is detected with APDs, as shown in Fig.~\ref{fig:APD_PNRD_comparison} up to $\lambda=0.8$.

\subsection{Lossy subtraction}\label{sec:APDloss}
\begin{figure*}
\centering
\subfigure[]{\includegraphics[width=0.32\textwidth]{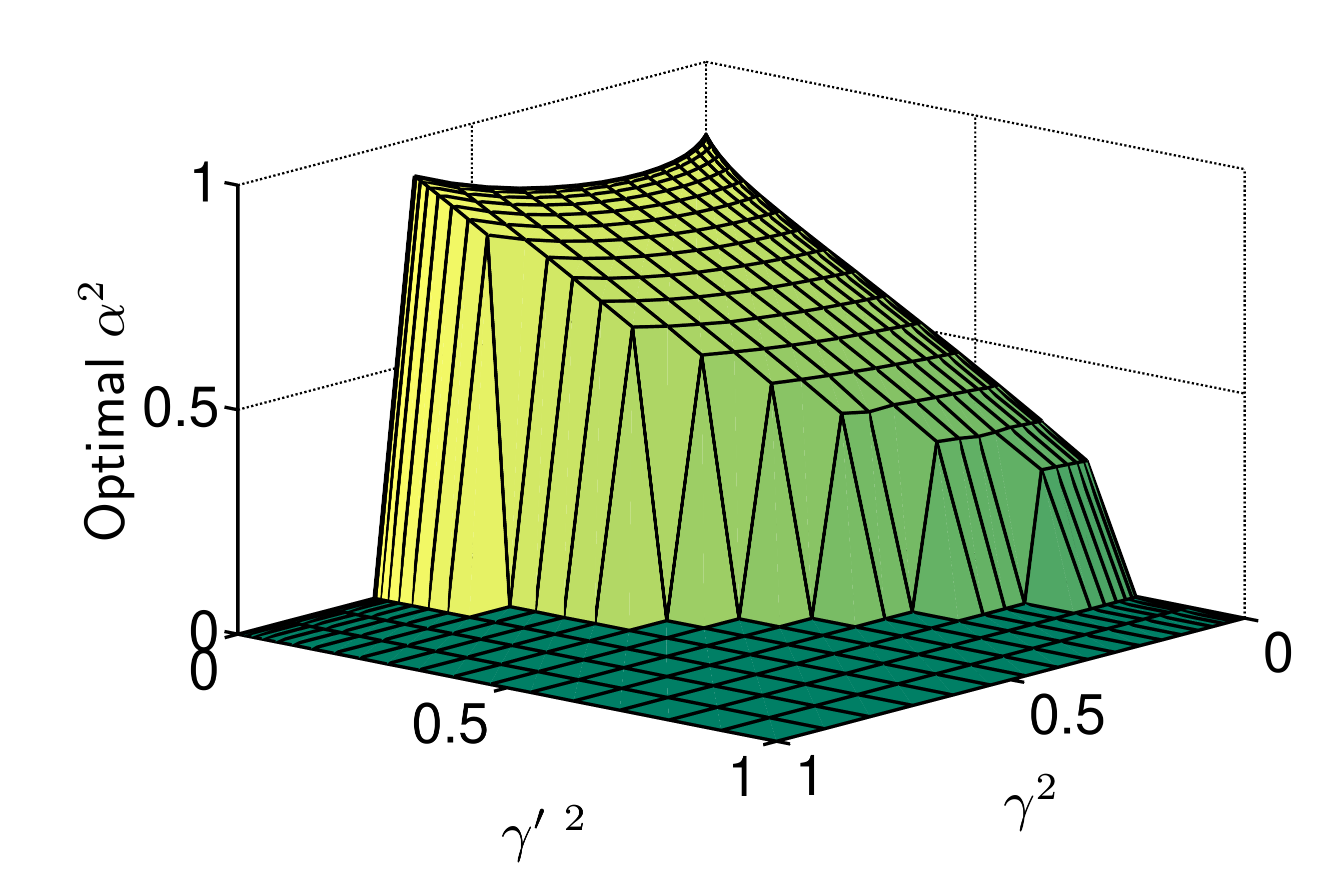}\label{fig:APDononA1}}
\subfigure[]{\includegraphics[width=0.32\textwidth]{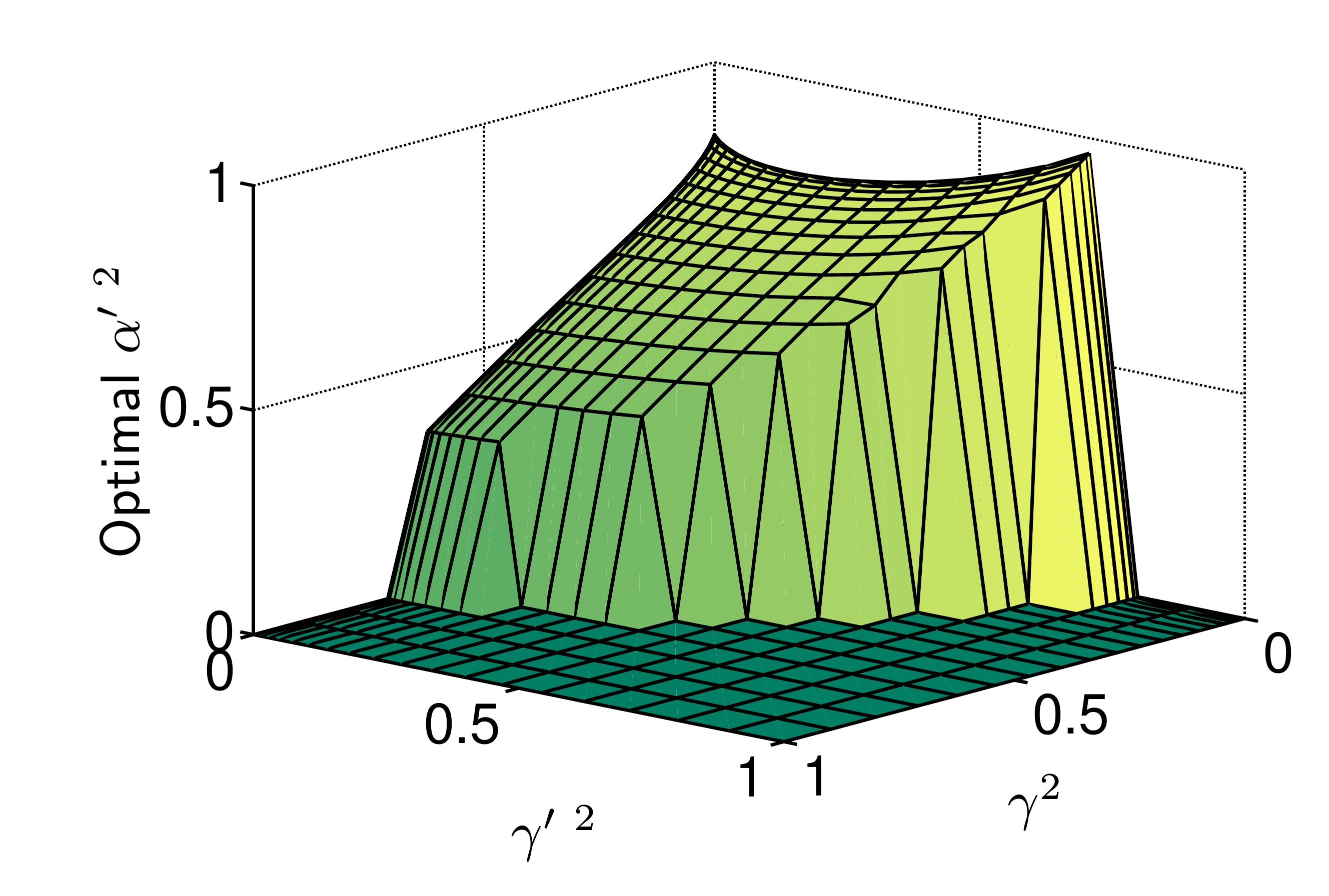}\label{fig:APDononA2}}
\subfigure[]{\includegraphics[width=0.32\textwidth]{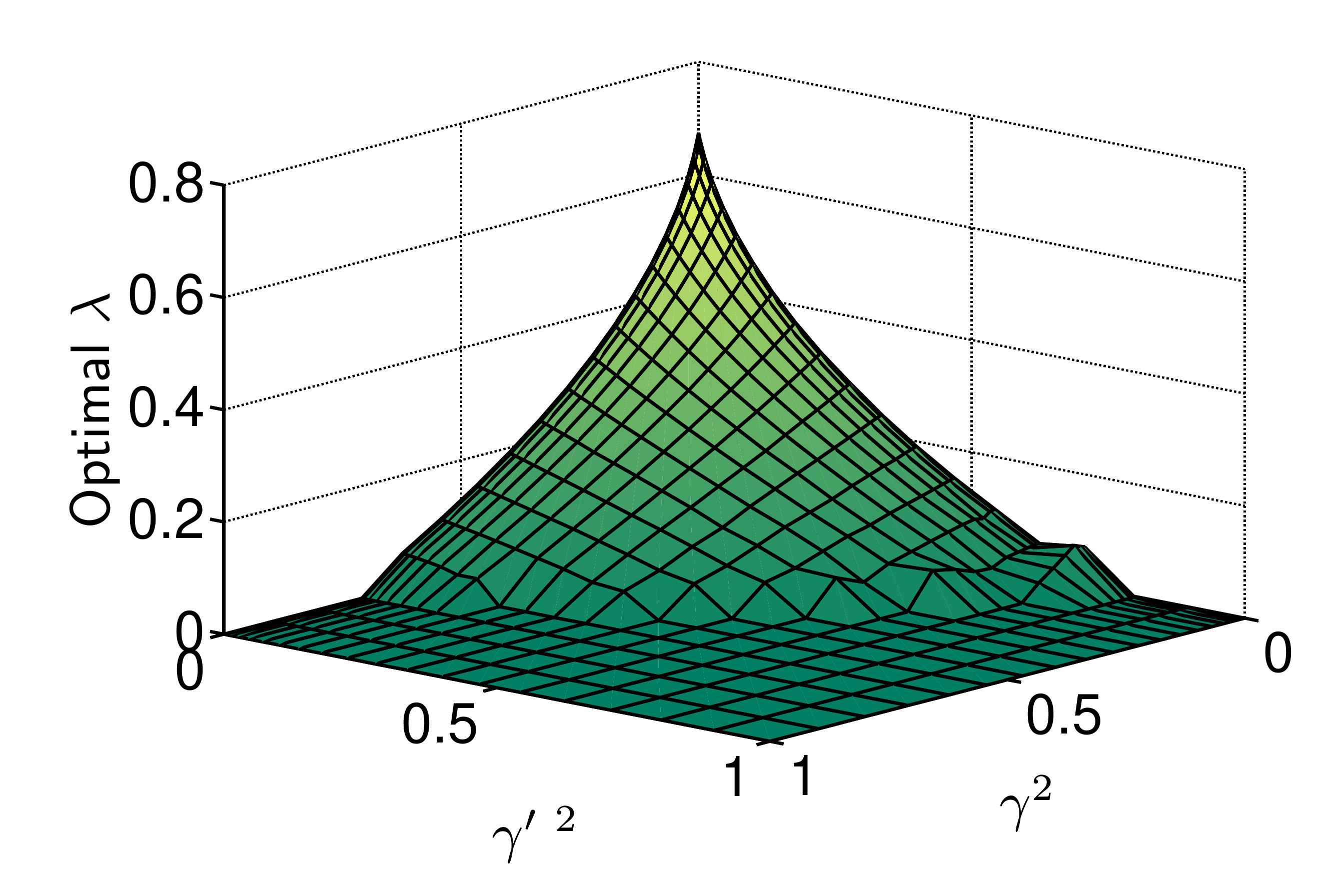}\label{fig:APDononLambda}}
\caption{(Color online) Lossy subtraction with APDs. \subref{fig:APDononA1}, \subref{fig:APDononA2}: subtraction parameters $\alpha^2$, $\alpha^{\prime2}$, respectively, and~\subref{fig:APDononLambda}: squeezing parameter $\lambda$, optimised to produce the highest entanglement gain rate under loss given symmetric subtraction with APDs.}\label{fig:APDononparams}
\end{figure*}

As with the PNRD case calculated above, we can use entanglement gain rate as the figure of merit to be optimised, from which the optimal parameters can be extracted. The optimal parameters $\alpha^2_\textrm{opt}$ and $\lambda_\textrm{opt}$ are shown in Fig.~\ref{fig:APDononparams}, with the resulting gain, probability and rate identical to that shown for PNRDs in Fig.~\ref{fig:PNRDAsymm}, as a function of the asymmetric loss parameters $\gamma^2,\gamma^{\prime2}$. Note that the optimal parameters are also the same for both the PNRD and APD cases.

\begin{figure*}
\centering
\subfigure[]{\includegraphics[width=0.3\textwidth]{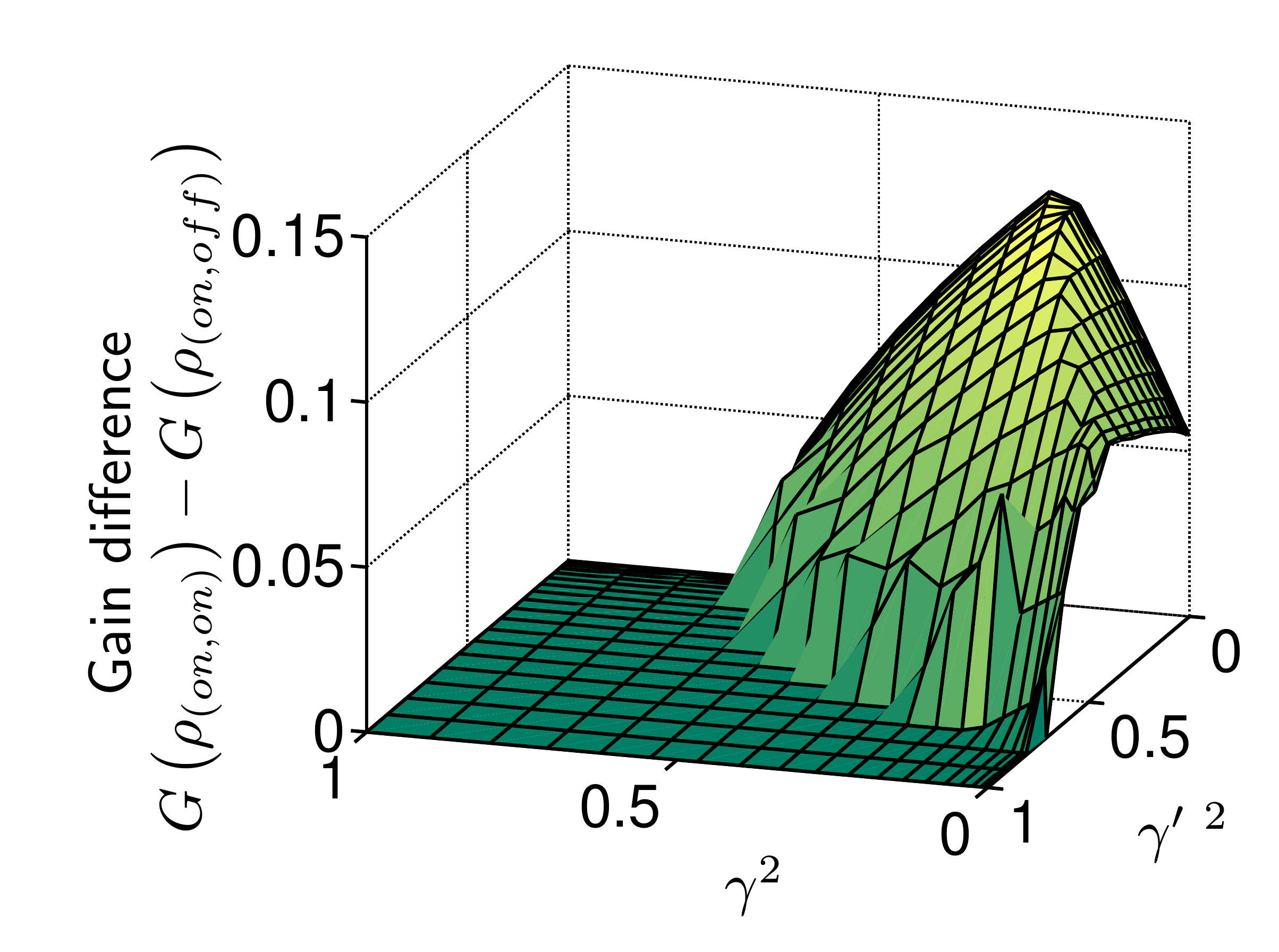}\label{fig:APDonoffGain}}\qquad
\subfigure[]{\includegraphics[width=0.3\textwidth]{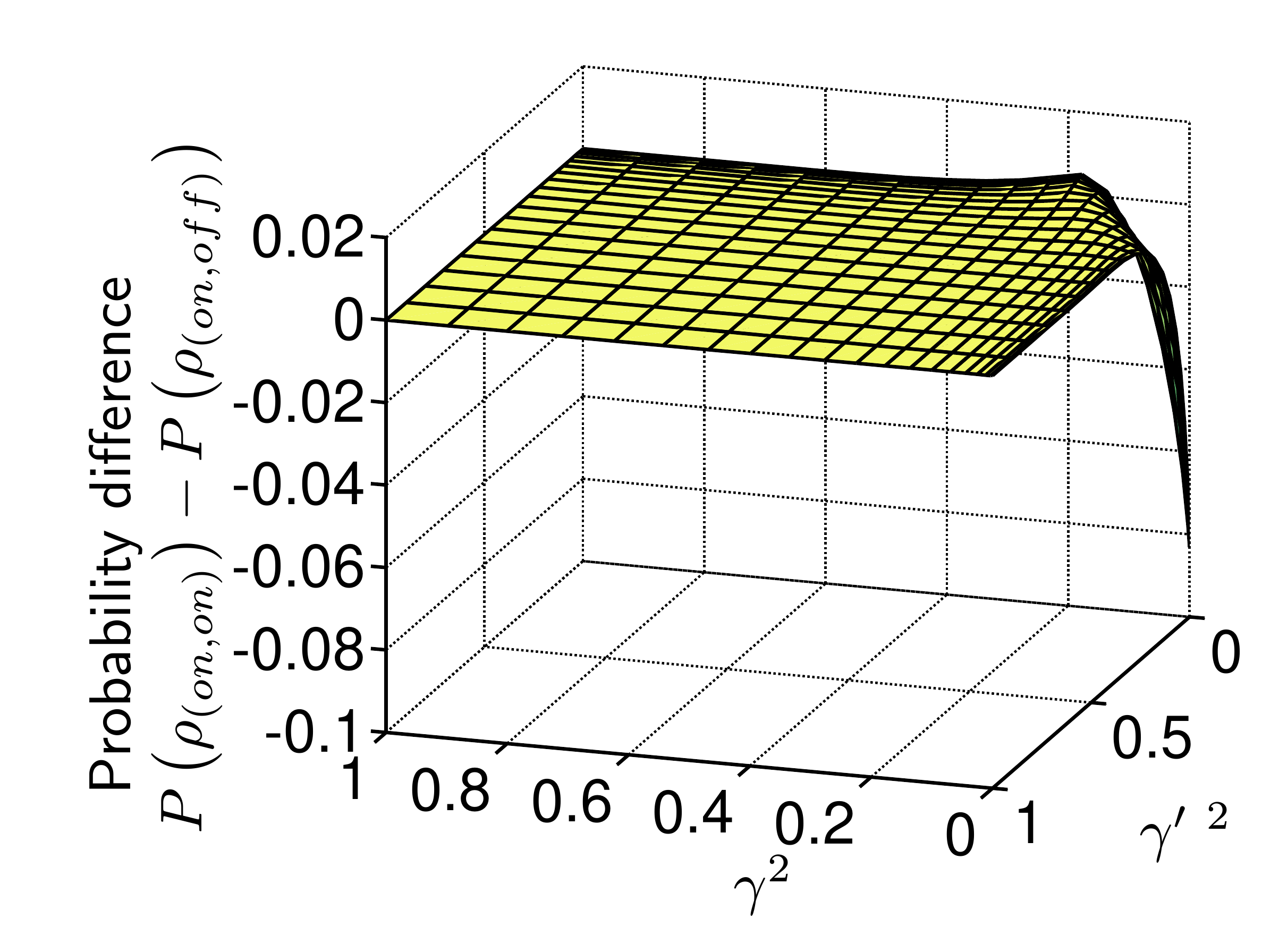}\label{fig:APDonoffProb}}\qquad
\subfigure[]{\includegraphics[width=0.3\textwidth]{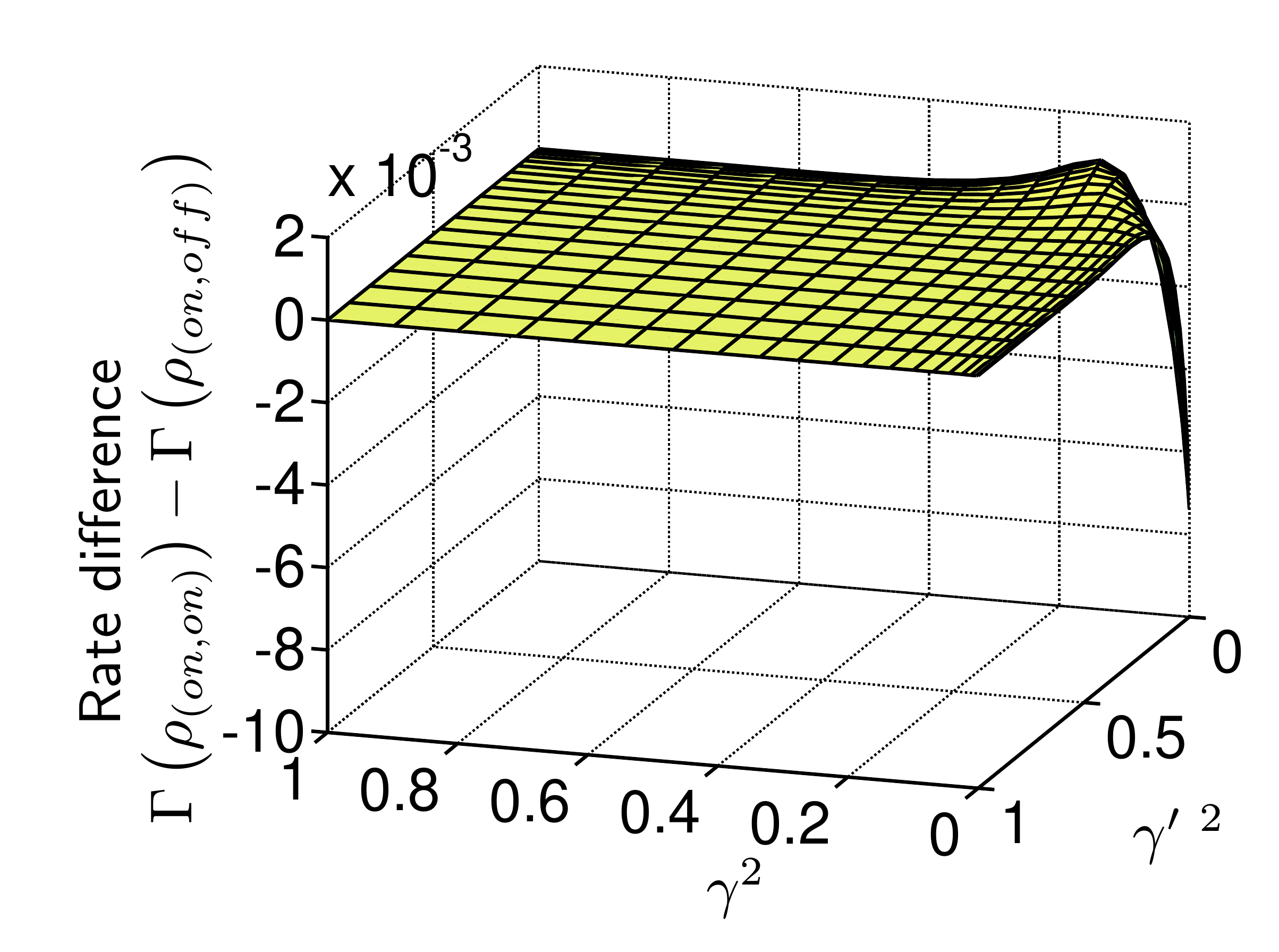}\label{fig:APDonoffRate}}
\caption{(Color online) Comparing symmetric and asymmetric subtraction with APDs under loss. The difference between (on,on) and (on,off) subtraction events in \subref{fig:APDonoffGain}~gain, \subref{fig:APDonoffProb}~probability and \subref{fig:APDonoffRate}~rate as a function of the losses $\gamma^2,\gamma^{\prime 2}$ on modes one and two respectively.}\label{fig:APDonoffRates}
\end{figure*}


\subsubsection{Imperfect Asymmetric Detection}\label{sec:APDasymmloss}
As with perfect subtraction, the behaviour of APDs and PNRDs under loss is the same at experimentally accessible values of $\lambda$. In the lossless case, we have shown that asymmetric subtraction always produces a higher rate of entanglement gain, (Sec.~\ref{sec:PNRDAsymm}, Fig.~\ref{fig:PNRDAsymm}). We now ask whether one strategy is always better than the other for all losses. This turns out not to be the case, as shown in Fig.~\ref{fig:APDonoffRates}.

While the gain produced when subtracting from both modes is always greater than when subtracting from a single mode (Fig.~\ref{fig:APDonoffGain}), for a particular range of losses on each mode $\gamma,\gamma^\prime$, the probability favours subtracting from just a single mode. This result is repeated in the plot showing the comparative rate of gain (Fig.~\ref{fig:APDonoffRate}).

\section{Conclusion}\label{sec:conc}
We have investigated photon subtraction from a two-mode squeezed state as a means to probabilistically increase entanglement under LOCC. By defining the entanglement gain rate as our figure of merit, we are able to optimise over the subtraction beam splitter transmissivity $T_2$ to maximise this quantity, in the presence of unequal losses on each mode. Our results may be summarised as follows:
\begin{enumerate}
\item If the losses are above a threshold, which depends on the number of photons to be subtracted, local subtraction cannot enhance entanglement.
\item When it can, (1,0) subtraction seems to be the best strategy, whether APDs or PNRDs are used to detect the subtracted photon.
\item However, depending on how losses are distributed across the modes, symmetric subtraction may be advantageous.
\item Subtracting more photons produces marginal enhancement in entanglement and is far less probable.
\item APDs are essentially equivalent to PNRDs for most of the $\lambda$ regime, including when accounting for asymmetric losses.
\end{enumerate}

Given any realistic scenario with lossy transmission channels and imperfect detectors, our conclusions outline the most suitable strategy that must be adopted to achieve entanglement enhancement in CV systems most effectively.  This approach also specifies the initial squeezing parameter $\lambda_\textrm{opt}$ for which this particular protocol works best. The figure of merit we use in this paper is the entanglement gain rate, relevant to quantum communication applications. For other applications, it may be the final entanglement, or indeed the gain alone, that is more important. The methods presented in this paper can be easily modified for those purposes. 

Photon subtraction is one of the simplest operations introducing non-Gaussianity, thereby opening the gate to a large class of CV quantum information processing protocols such as entanglement distillation that are not possible in the Gaussian regime. We hope that our work will inform future efforts to increase entanglement via this technique under realistic experimental conditions.

\section*{Acknowledgements}
The authors are grateful to M. Barbieri for useful discussions and P. van Loock for helpful correspondence. This work was supported by the Engineering and Physical Sciences Research Council of the United Kingdom (Project No. EP/H03031X/1), the US European Office of Aerospace Research \& Development (Project No. 093020), the European Commission (under Integrated Project
Quantum Interfaces, Sensors, and Communication based on
Entanglement) and the Royal Society. 

\bibliography{references3}

\end{document}